\documentclass[preprint,preprintnumbers,amsmath,amssymb,floatfix,nofootinbib,superscriptaddress]{revtex4}
\usepackage[latin9]{inputenc}
\usepackage{amssymb}
\usepackage{graphicx}

\makeatletter

\@ifundefined{textcolor}{}
{%
 \definecolor{BLACK}{gray}{0}
 \definecolor{WHITE}{gray}{1}
 \definecolor{RED}{rgb}{1,0,0}
 \definecolor{GREEN}{rgb}{0,1,0}
 \definecolor{BLUE}{rgb}{0,0,1}
 \definecolor{CYAN}{cmyk}{1,0,0,0}
 \definecolor{MAGENTA}{cmyk}{0,1,0,0}
 \definecolor{YELLOW}{cmyk}{0,0,1,0}
}

\makeatother

\begin{document}


\title{Exploring the Hadronic Axion Window via Delayed Neutralino Decay to
Axinos at the LHC}

\author{C.S.~Redino}

\email{csredino@buffalo.edu}

\affiliation{Department of Physics, SUNY at Buffalo, Buffalo, NY 14260-1500, USA}

\author{D.~Wackeroth}

\email{dow@ubpheno.physics.buffalo.edu}

\affiliation{Department of Physics, SUNY at Buffalo, Buffalo, NY 14260-1500, USA}

\date{\today}

\begin{abstract}
The addition of the QCD axion to the Minimal Supersymmetric Standard Model (MSSM)
not only solves the strong CP problem but also modifies the dark sector with
new dark matter candidates. While SUSY axion phenomenology is usually
restricted to searches for the axion itself or searches for the ordinary
SUSY particles, this work focuses on scenarios where the axion's superpartner,
the axino, may be detectable at the Large Hadron Collider (LHC) in
the decays of neutralinos displaced from the primary vertex. In particular,
we focus on the KSVZ axino within the hadronic axion window.
The decay length of neutralinos in this scenario easily fits the ATLAS
detector for SUSY spectra expected to be testable at the 14 TeV LHC.
We compare this signature of displaced decays to axinos to other
well motivated scenarios containing a long lived neutralino which
decays inside the detector. These alternative scenarios can in some
cases very closely mimic the expected axino signature, and the degree
to which they are distinguishable is discussed. We also briefly comment
on the cosmological viability of such a scenario.
\end{abstract}

\maketitle

\section{Introduction}

\label{sec:intro}

Supersymmetry (SUSY)~\cite{Wess:1973kz,Wess:1974tw} is a
well-motivated framework for physics beyond the Standard Model (SM)
near the weak scale (see, e.g.,
\cite{Fayet:1976cr,Nilles:1983ge,Haber:1984rc,Martin:1997ns,Signer:2009dx}
for a review), with attractive features such as a solution to the
weak-scale gauge hierarchy problem in the SM, the possibility of gauge
coupling unification, and an apparent dark matter (DM) solution. The
lightest neutralino in the Minimal Supersymmetric SM (MSSM) acting as
a weakly interacting massive particle (WIMP) provides approximately
the correct cold DM (CDM) relic abundance (for a review see,
e.g.,~\cite{Jungman:1995df}). This so-called ``WIMP miracle'' in SUSY
is often overstated, however, in that there can still exist a tuning
of SUSY parameters in order to get the correct abundance, SUSY models
with the correct abundance only constitute a very small fraction of
SUSY model space, and models that over or under predict the DM relic
abundance are
common~\cite{Hooper:2013qjx,Huang:2014xua,Baer:2012mv,Baer:2014eja,Amsel:2011at}.
Simply by virtue of reducing the available parameter space, the
requisite of a DM solution can exclude more natural scenarios in SUSY
\cite{Baer:2012mv}.

The additional constraints and tuning introduced by accommodating DM
in a SUSY model can be avoided by extending the dark sector beyond
just neutralinos, but it is desirable to do this in a way that is
minimal and well motivated in of itself. The neutral, pseudoscalar,
R-parity even axion is an attractive candidate for an extended dark
sector because it has motivation beyond its properties as a DM
particle. The Peccei-Quinn (PQ) axion's origins are independent of any
considerations of DM, originally introduced to resolve an apparent
fine-tuning problem in the SM QCD sector, known as the strong CP
problem~\cite{Peccei:1979,Peccei:2006as}. The strength of the axion's
interactions is suppressed by the PQ scale $f_{a}$. When
$f_{a}$ is sufficiently large the axion becomes a viable DM candidate~\cite{Abbott:1982af,Preskill:1982cy,Dine:1982ah}.

The role of the axion as a DM candidate in a SUSY model can be more
complicated than simply complimenting the neutralino relic
abundance. Embedded in a SUSY model, the axion is a member of a
supermultiplet, joined by a neutral, R-parity odd Majorana chiral
fermion, the axino~\cite{Weinberg:1977ma,Wilczek:1977pj}, and the
R-parity even scalar saxion \cite{Kim:1983ia}. Also the axino is a
viable DM candidate~\cite{Rajagopal:1990yx} and can provide the
correct CDM relic
abundance~\cite{Covi:1999ty,Brandenburg:2004xh,Strumia:2010aa,Choi:2011yf}.
With the neutralino, axion and axino being all valid DM candidates
(see, e.g.,~\cite{Baer:2014eja} for a recent review), which particle
(or set of particles) actually plays the role of DM in a given model
depends on their mass hierarchy and the cosmology, so there are a
variety of possible scenarios. The scenarios in such a PQ-augmented
MSSM (PQMSSM) have been studied extensively
in~\cite{Baer:2009vr,Baer:2011hx,Bae:2013hma,Baer:2011uz,Baer:2010wm},
where it has been pointed out that there exist significantly more
scenarios than in the MSSM that predict the correct CDM abundance.
Apart from solving the strong CP problem, the PQMSSM can have other
benefits such as Yukawa unification \cite{Baer:2012cf} or being
embedded in a full GUT theory \cite{Baer:2012jp}. Moreover, scenarios
in the PQMSSM often have an easier time accommodating naturalness in
SUSY, if for nothing else then because an extended model has more
parameters and is more flexible.

These attractive features of a PQMSSM warrant an extensive study of
all aspects of its phenomenology.  While SUSY axion phenomenology is
usually restricted to searches for the axion itself or searches for
the ordinary SUSY particles, this work focuses on scenarios where the
axion's superpartner, the axino, may be detectable at the Large Hadron
Collider (LHC).  In particular, we focus on the
Kim-Shifman-Vainshtein-Zhakharov
(KSVZ)~\cite{Kim:1979if,Shifman:1979if} axion model, and consider the
axino to be the lightest SUSY particle (LSP) produced at the LHC in
neutralino decays displaced from the primary vertex.  The latter can
happen when the PQ scale $f_a$ is in the range known as the hadronic
axion window \cite{Chang:1993gm}. The decay length of neutralinos in
this scenario easily fits the ATLAS detector for SUSY spectra expected
to be testable at the 14 TeV LHC. We explore the possibility of
distinguishing this axino collider signature from other well motivated
scenarios containing a long lived neutralino but no axino
LSP. Examples of alternate collider signatures of axino LSPs, for
instance in displaced charged slepton decays and in prompt or
displaced Higgsino decays can be found in
respectively~\cite{Brandenburg:2005he} and \cite{Barenboim:2014kka}
(see Section~\ref{sec:background} for a more detailed discussion).  To
make the collider phenomenology of the model under consideration in
this work possible at all requires certain assumptions about the axion
model and the SUSY spectra, which makes this scenario distinct from
those already studied, but nonetheless it is a predictive scenario
with the possibility of low tuning, and compatibility with an
attractive cosmology.

The rest of the paper is organized as follows: In
Section~\ref{sec:background}, we discuss what assumptions are
necessary for an axion model so that collider phenomenology is
possible. In Section~\ref{sec:signal}, we discuss the proposed signal
and an example SUSY benchmark model with parameters that put this
signal within reach at the 14 TeV LHC. In Section~\ref{sec:results},
we compare this signal in detail to other similar possibilities from
gravitinos and R-parity violation (RPV). A discussion of how the
scenarios under study can accommodate the correct DM abundance can be
found in Section~\ref{sec:remarks}. Lastly we conclude by considering
the limitations of this work and how it can be expanded in the
future. A greatly expanded discussion of the proposed scenario and the
results in this study can be found in~\cite{Redino:2015rsa}.

\section{Motivation and Assumptions}

\label{sec:background}

\begin{figure}[t]
\centerline{\includegraphics[width=6.5in]{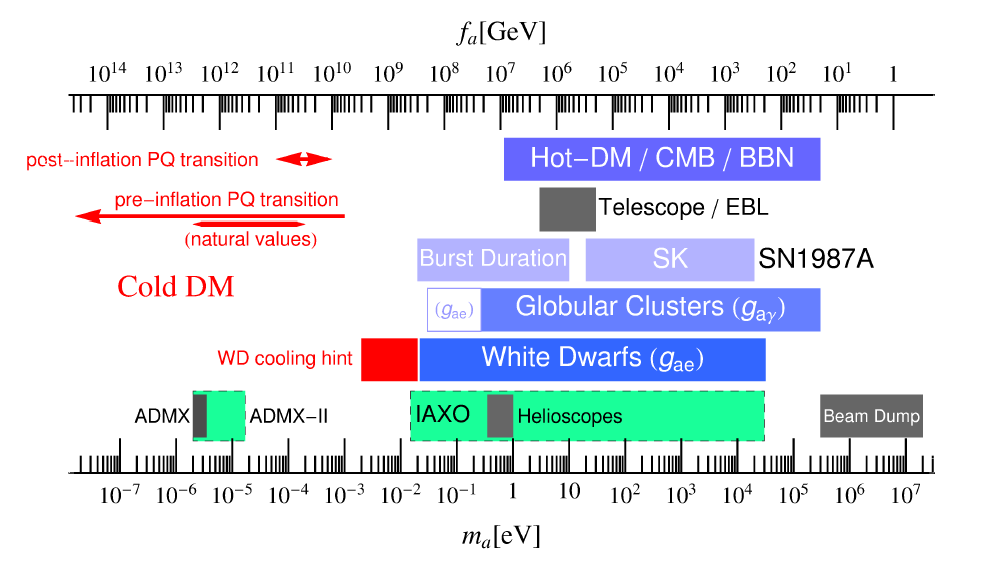}}
\caption{Exclusion ranges for the Peccei-Quinn scale $f_{a}$ (and the related axion mass $m_a$) from various
constraints as described in~\cite{Agashe:2014kda}. A critical discussion of these constraints in the scenario 
considered in this work can be found in the text. \label{fig:one}}
\end{figure}

The usual wisdom that prevents axinos from being considered for
collider phenomenology is their extremely weak coupling.  All the
couplings of the axion are suppressed by the PQ scale $f_{a}$, which
in theory can take any value. Axions as CDM candidates are usually
only considered with $10^{9}\mbox{
  GeV}<f_{a}<10^{14}$~GeV~\cite{Kim:2008hd}.  For an axion to be the
Pseudo-Nambu-Goldstone boson of a spontaneously broken global $U(1)$
chiral symmetry to solve the strong CP problem, the axion field
($a(x)$) described as follows~\cite{Peccei:1979} (for a review see,
e.g.,~\cite{Peccei:2006as}):
\begin{equation}
\label{eq:pqaxion}
{\cal L}_{axion} = -\frac{1}{2} \partial_\mu a(x) \partial^\mu a(x)+{\cal L}_{int}(\partial^\mu a/f_a;\Psi))+ \xi\frac{a(x)}{f_{a}}\frac{\alpha_s}{8\pi}F_{b}^{\mu\nu}\tilde{F}_{b\mu\nu} \, ,
\end{equation}
has an effective potential that has a minimum at $<a>=-\frac{f_a}{\xi}
\bar \theta$. $F_b^{\mu\nu}$ is the gluon field strength tensor, and
$\tilde F^{\mu\nu}_b$ its dual. Thus, with $a_{phys}=a-<a>$ the
CP-violating $\bar \theta$-vacuum term in ${\cal L}_{QCD}$ is
canceled. As can be seen in Eq.~\ref{eq:pqaxion}, the coupling
strength of the axion to the SM particles is governed by $f_a$ with
$f_{a}$ being a free model parameter, only determined by experimental
and observational constraints.  A summary of all the current
constraints on $f_{a}$ is shown in
Fig.~\ref{fig:one}~\cite{Agashe:2014kda}.

The model-dependent nature of these limits can be exploited to
conceive models appropriate for collider phenomenology, but first it
is worth considering why the typical limits are so restrictive. If the
usual lower bound of $f_{a}>10^{9}$~GeV is taken at face value then
collider studies for axions (or equally suppressed axinos) are very
limited.  Direct production rates of axions/axinos are simply too
small with such a suppressed coupling, but an incredibly weak coupling
may still be probed in certain cases, usually by taking advantage of
R-parity.  If an extremely weakly coupled particle (such as an axino
or gravitino) is the lightest particle with an odd R-parity then their
appearance at the end of a SUSY decay chain is inevitable, regardless
of the coupling size. One way to take advantage of this is if the
next-to-lightest SUSY particle (NLSP) is charged, then its delayed
decay to the suppressed LSP will leave a track (see, e.g.,
Ref.~\cite{Brandenburg:2005he}).  Even with neutral NLSPs, R-parity
can still be exploited in much the same way to look for extremely
weakly coupled particles, but with a less spectacular signal. R-parity
still requires the LSP to be at the end of any SUSY decay chain, and
while there is no longer a charged track, the visible decay products
in the last leg will be displaced. Even though R-parity forces a
branching fraction of one for the LSP, the very weak coupling of the
LSP still has an effect: in determining the width of the NLSP which
translates to how displaced the last leg of the SUSY cascade will
be. If the suppression factor is great enough, then the displaced
decay occurs completely outside the detector, and so the scenario is
indistinguishable from one without the extra particle in the final leg
at all. If these types of searches are taken seriously for gravitinos
(and they are \cite{Chatrchyan:2012ir,Aad:2014gfa}), one would hope
that this could be exploited for axinos, but there are a couple of
technical differences that make this difficult. Naively one would
assume these searches are even harder for gravitinos since their
interactions are suppressed by the Planck scale, which dwarfs even the
higher values of $f_{a}$ that are considered in the literature.  While
the gravitino's interactions are suppressed by the Planck scale, they
are also inversely proportional to the gravitino mass and so the
suppression is not as great as one may naively
expect~\cite{Ambrosanio:1996jn}.  In effect, the coupling strength can
be tuned to any value provided there is freedom in choice of the
gravitino mass (which can take a wide range of values depending on the
model). For extremely small gravitino masses, searches for displaced
or even prompt decays of NLSPs become possible. In displaced decays to
axinos, the effective coupling is relatively insensitive to the axino
mass and only depends strongly on $f_{a}$ and any axino with
$f_{a}>10^{9}$~GeV is expected to be completely invisible at
colliders, as any decaying NLSP will always leave the detector
\cite{covi:11:moriondEW}. Very recently an exception to this common
wisdom was explored in~\cite{Barenboim:2014kka}, where the authors
showed that a Higgsino NLSP can decay to an axino LSP inside the
detector easily even with $f_{a}>10^{9}$~GeV, when there is a direct
coupling between Higgsinos and axinos (and an appropriate mass
spectrum) as in the case of Dine-Fischler-Srednicki-Zhitnitsky (DFSZ)
axions~\cite{Dine:1981rt,Zhitnitsky:1980tq}.

For DFSZ axions, displaced decays of a neutral NLSP are possible because
of the coupling between axinos and Higgs bosons/Higgsinos, but for
the other main class of axion model, the KSVZ axion, it seems that
collider studies are only possible, if the constraints on $f_{a}$
can be relaxed. 

Limit plots, such as Figure~\ref{fig:one}, where all the constraints
are given in terms of $f_{a}$, often do not easily reveal the
underlying model-dependent assumptions in extracting these
constraints. After closer inspection, it turns out that in the KSVZ
model there is the intriguing possibility of evading most of the
constraints in a way not possible in the DFSZ model. In the KSVZ model
the axion coupling to leptons is vanishing at tree level, and the
effective coupling to leptons at one loop has been shown to be
non-constraining \cite{Srednicki:1985xd,Kaplan:1985dv}.  As a result,
a whole category of constraints, i.e. the limits from white dwarf
cooling, are irrelevant for the KSVZ axion when $g_{ae}=0$ in
Fig.~\ref{fig:one}. In the DFSZ model, the coupling to leptons is
always non-vanishing. Aside from the white dwarf limits, the coupling
most often tested is the axion coupling to photons, given by
\begin{equation}
{\cal L}_{a\gamma\gamma}=\frac{\alpha}{4\pi}K_{a\gamma\gamma}\frac{a_{phys}}{f_{a}}F^{\mu\nu}\tilde{F}_{\mu\nu}\;.
\end{equation}
where $F^{\mu\nu}$ is the photon field strength tensor and $K_{a\gamma\gamma}$
parameterizes the model-dependent axion-photon coupling. This too
can be made to vanish in the KSVZ model (but not the DFSZ). In the KSVZ model
the coupling to photons is determined by~\cite{Kaplan:1985dv,Srednicki:1985xd} 
\begin{equation}
K_{a\gamma\gamma}=K_{a\gamma\gamma}^{0}-\frac{2(4+z)}{3(1+z)}\;,\label{eq:kgg}
\end{equation}
and the difference can be thought of as a balancing between short and
long-range effects. The short-range effect $K_{a\gamma\gamma}^{0}$
comes from the chiral anomaly depending on the electromagnetic
charge(s) of the new heavy quark(s) that are required for UV
completion in KSVZ models. The second term in Eq.~\ref{eq:kgg} comes
from the axion's mixing with light mesons and depends on the value of
$z$, the ratio of light quark masses ($z=m_u/m_d$), which comes with
some uncertainty ($z=0.38-0.58$~\cite{Agashe:2014kda}). For an
appropriate choice of charge(s) for the new quark(s) these two terms
can cancel and $K_{a\gamma\gamma}$ can be made to vanish
\cite{Kaplan:1985dv,Srednicki:1985xd}. It should be noted that to
avoid the existing constraints $K_{a\gamma\gamma}$ does not have to
vanish exactly, but only be so small that it is consistent with the
limits on the photon coupling. Once $g_{a\gamma}=0$ is assumed the
constraints originating from globular clusters and from telescope
searches for $a\to \gamma\gamma$ in Fig.~\ref{fig:one} are no longer
valid.  Also current and anticipated constraints from the search for
solar axions at CAST~\cite{Arik:2015cjv}, and
IAXO~\cite{Armengaud:2014gea} rely on a non-zero axion-photon
coupling.

Thus, for a KSVZ axion with no photon or lepton coupling ($g_{ae}=0$
and $g_{a\gamma}=0$ in Fig.~\ref{fig:one}), the only
constraints in Figure~\ref{fig:one} that are truly inescapable are
those coming from SuperNova (SN) 1987a
\cite{Turner:1989vc,Raffelt:1990yz}.  In practice, the new light
species do not even have to be detected from these events, but their
properties can be constrained from the burst duration of the
supernova, and the number of particles detected from the light species
of the Standard Model (neutrinos). In the case of QCD axions this is
especially interesting because it directly tests the otherwise elusive
gluon-axion coupling, the only coupling necessary to solve the strong
CP problem, and the only coupling free from model dependent
factors. As seen in Figure~\ref{fig:one}, there are actually two
separate regions of bounds from SN1987a, corresponding to two regimes
of coupling strength. The upper bound range comes from when the axion
is so weakly coupled that it is free streaming after it is produced in
the supernova and the lower bound range is from axions that still have
interactions with nuclear material on there way out of the supernova.
If the photon coupling is taken to vanish (or be adequately
suppressed), then there is a gap in the constraints between the two
regimes of free streaming and interacting axions in Supernovae. This
window allows a range of lower suppression scale $3\times10^{5}{\rm
  GeV}<f_{a}<3\times10^{6}{\rm GeV}$, known as the hadronic axion
window, which has been examined in the literature in the past,
particularly in the context of axions as hot dark matter and its
cosmological implications~\cite{Moroi:1998qs,Chang:1993gm}.  Though
hot dark matter is now greatly disfavored, axions in this window,
though still relatively suppressed by $f_{a}$, should have coupling
strengths such that their partners, the axinos, can be studied at
colliders, via the general strategy for gravitinos and displaced
decays described above. The hot dark matter bound in
Figure~\ref{fig:one} is the usual reason why the hadronic axion window
is considered ``ruled out'', which comes about from there being too
much of a hot thermal relic of axions in conflict with measurements of
large scale structure and the CMB, but at least one way to avoid this
is by considering cosmologies with a very low reheating
temperature. Although the hot dark matter bound is model independent
with regards to the axions themselves, it is model dependent as far as
the cosmology is concerned. A dangerous hot thermal relic in one
cosmology can be made safe in another cosmology with a lower reheating
temperature. Standard Model neutrinos were shown to be a viable warm
dark matter candidates with this method in \cite{Giudice:2000dp} and
the same principle has been applied to axions to alleviate constraints
\cite{Grin:2007yg}. By lowering the reheating temperature sufficiently,
thermal relics freeze out while the universe is still undergoing
inflation, and the relic abundance will be diluted. Diluting the
thermal axion abundance sufficiently in a scenario where $f_{a}$ is in
the hadronic axion window would evade constraints,
but could also require an additional species of dark matter,
such as an axino.

With these assumptions laid out, the scenario to be studied at the LHC
should be clear: KSVZ axinos with a neutral NLSP, with only a QCD
coupling, and the suppression scale, $f_{a}$, to be considered lying
in the range given by the hadronic axion window:
\[
3\times10^{5}\,{\rm GeV}<f_{a}<3\times10^{6}\,{\rm GeV}
\]
This scenario is motivated by being perhaps the only KSVZ model that
is testable at a collider without a charged NLSP. This scenario also
has the possibility of having low tuning and an interesting cosmology
with its own testable consequences. In Section~\ref{sec:remarks}, we
will provide an estimate of the DM abundance due to
thermal/non-thermal axions and axinos in this model.

\section{Signal and Benchmark}

\label{sec:signal}

In the following we will concentrate on PQMSSM scenarios with
a neutralino NLSP. With only the effective coupling to gluons being allowed in the
hadronic axion window, this makes for a very predictive scenario.

\begin{figure}
\centerline{\includegraphics[scale=0.3]{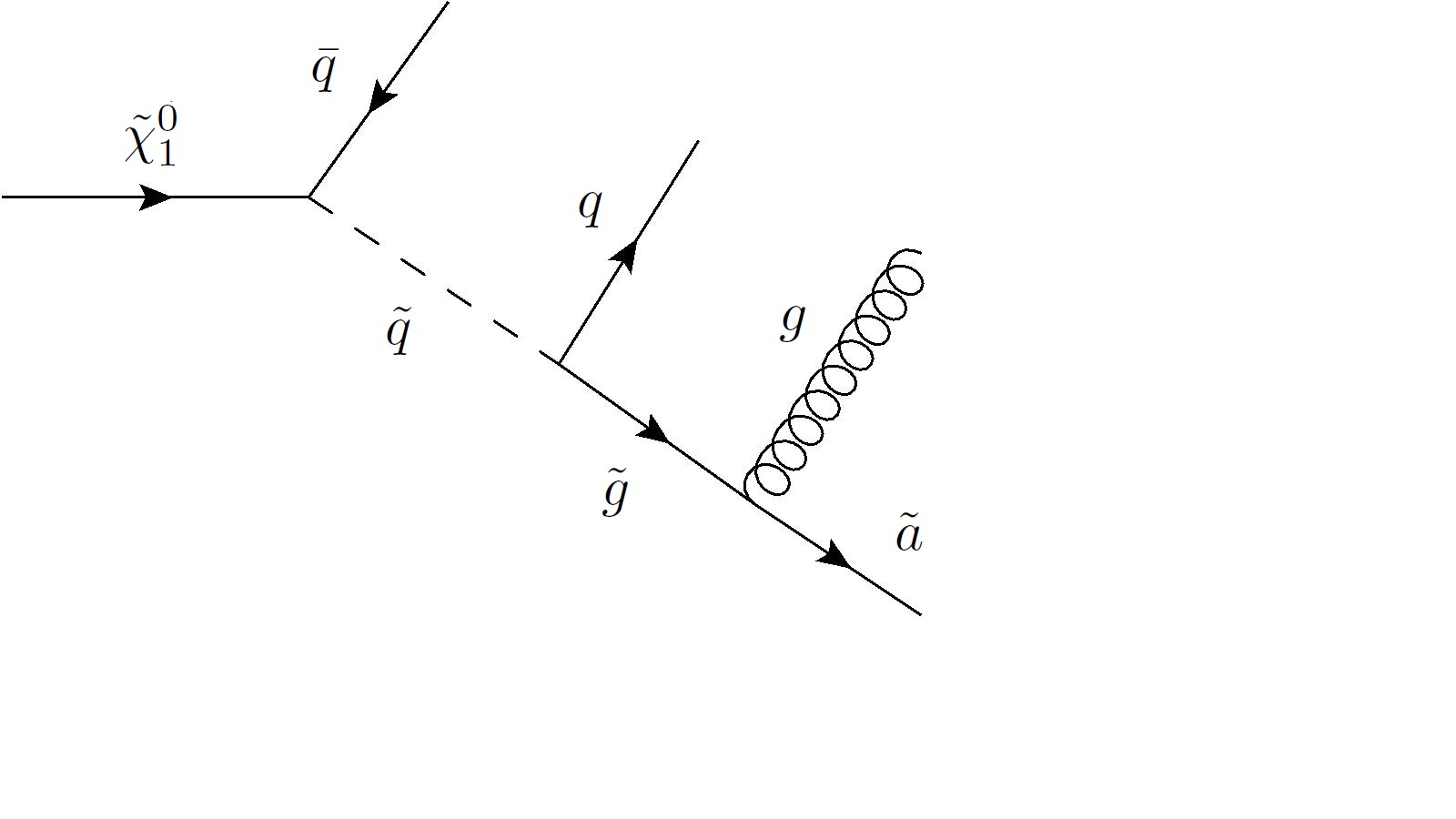}} \caption{Neutralino decay to three jets and an axino via ${\cal L}_{\tilde{a}\tilde{g}g}$ of Eq.~\ref{eq:agg}.\label{fig:two}}
\end{figure}

The supersymmetric version of the axion-gluon coupling is the axino-gluino-gluon
coupling~\cite{Baer:2011hx}, 
\begin{equation}
{\cal L}_{\tilde{a}\tilde{g}g}=i\frac{\alpha_{s}}{16\pi f_{a}}\bar{\tilde{a}}\gamma_{5}[\gamma^{\mu},\gamma^{\nu}]\tilde{g}_{b}F_{b\mu\nu}\label{eq:agg}
\end{equation}
and is the only coupling available to produce axinos in this scenario.
$\tilde{a}$ and $\tilde{g}_{b}$ denote the axino and gluino field
respectively. Even within the window of lower $f_{a}$ considered here, the suppression
is still too large to expect production of axinos at the LHC unless
they follow the NLSP in a decay chain so that there are no other less
suppressed options for decay. Once a neutralino is produced there
is only one dominant topology for its decay to an axino at tree level
(Fig.~\ref{fig:two}), via an off shell squark and an off shell gluino,
resulting in missing transverse energy (MET) and three displaced jets,
(plus whatever SM particles were produced in association with the
neutralino). This topology allows decays to heavy quarks, but the
decay width should be relatively small compared to that of decays
to light quarks, provided the neutralino is not too massive. At tree
level this is the only topology that leads to four decay products
and there are no topologies with a smaller multiplicity. Any other
decay path from neutralino to axino involves more final state particles
and possibly more massive off shell sparticles in the decay chain,
and so the process is even more greatly suppressed to the point where
it is negligible compared to the three displaced jets and MET channel. 

\begin{figure}
\centerline{\includegraphics[scale=0.3]{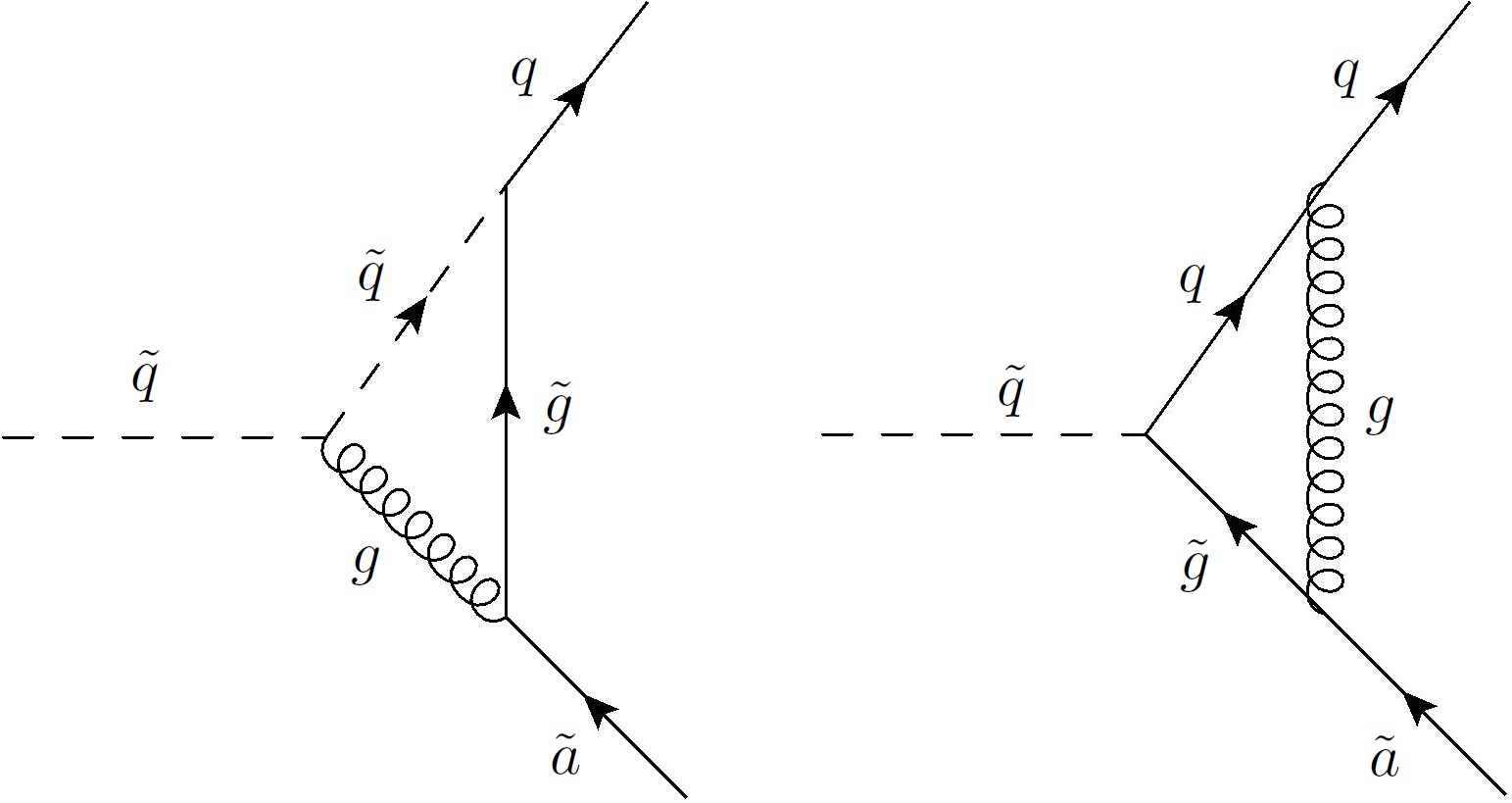}}
\caption{The vertex corrections that lead to the effective squark-quark-axino
interaction, ${\cal L}_{\tilde{a}q\tilde{q}}$ of Eq.~\ref{eq:aqq}.\label{fig:three}}
\end{figure}

\begin{figure}
\centerline{\includegraphics[scale=0.3]{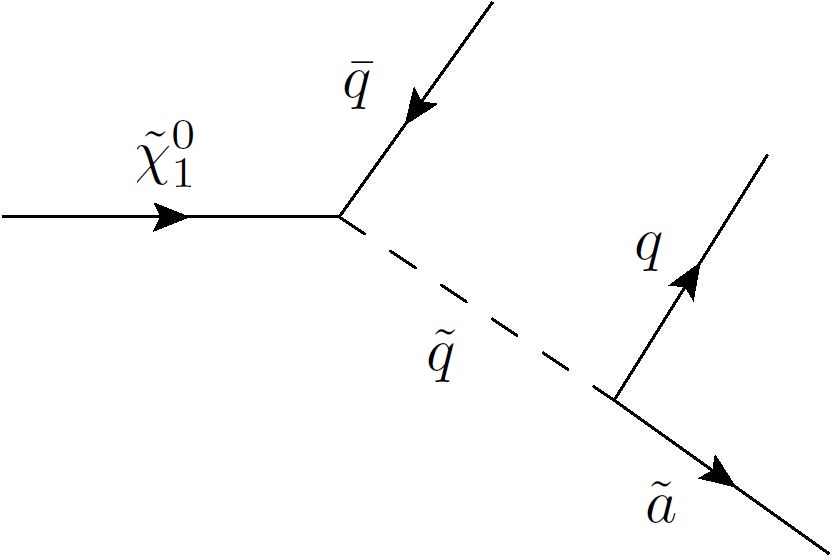}} \caption{Neutralino decay to two jets and an axino via ${\cal L}_{\tilde{a}q\tilde{q}}.$ of Eq.~\ref{eq:aqq}. \label{fig:four}}
\end{figure}

It is very important however to consider one-loop effects here. The
vertex correction to squarks decaying to axinos, shown in Figure~\ref{fig:three},
provides an effective squark-quark-axino coupling that can provide
the dominant decay channel for neutralinos for large swathes of SUSY
parameter space. This effective coupling was first explored in \cite{Covi:2002vw},
and with the heavy states integrated out, this interaction takes the
form 
\begin{equation}
{\cal L}_{\tilde{a}q\tilde{q}}=-g_{eff}\tilde{q}_{j}^{L/R}\bar{q}_{j}P_{R/L}\tilde{a}\;,\label{eq:aqq}
\end{equation}
where $m_{\tilde{g}}$ is the gluino mass, $\bar{q}_{j}$ and $\tilde{q}_{j}^{L/R}$
is the quark and (left or right-handed) squark field respectively,
and $P_{R/L}=(1\pm\gamma_{5})/2$, and the effective coupling 
\begin{equation}
g_{eff}\simeq \frac{\alpha_{s}^{2}}{\sqrt{2}\pi^{2}}\frac{m_{\tilde{g}}}{f_{a}}\log\left(\frac{f_{a}}{m_{\tilde{g}}}\right)\;.\label{eq:geff}
\end{equation}
With this effective coupling considered, there is the possibility
of neutralino decay to an axino and two jets (Figure~\ref{fig:four}),
and this is the decay channel we focus on here. The relative strength
of ${\cal L}_{\tilde{a}\tilde{g}g}$ and ${\cal L}_{\tilde{a}q\tilde{q}}$
was explored in \cite{Covi:2002vw} with regards to squark decays,
where it was shown that the ${\cal L}_{\tilde{a}q\tilde{q}}$ decay
dominates unless $m_{\tilde{q}}\gg m_{\tilde{g}}$. This also holds
true here, where neutralino decays are mediated by an off-shell squark.
The decay width for $\tilde{\chi}_{j}^{(0)}\to q\bar{q}\tilde{a}$
is discussed in more detail in Section~\ref{sec:width}.

For two produced neutralinos decaying to axinos, the signal is always
multi-jets and MET, and depending on the SUSY spectra the dominate
decay will either be two or three jets per decay leg (before showering/clustering).
Multi-jets and MET is by far the most commonly studied signal for
prospective new physics at the LHC, but if the jets are displaced
enough, then the signal for KSVZ axinos with a neutralino NLSP can
become rather unique. If the jets are displaced enough when neutralinos
decay then the SM background will become negligible, and the only
competing or alternative source for such a signal would come from
other new physics. Two such alternative sources are displaced decays
of neutralinos to gravitinos and displaced decays of neutralinos to
neutrinos via RPV. These two alternative sources may not only arise
in alternative models, but could all exist consistently in one model,
i.e, a model with axinos, light gravitinos and RPV couplings all at
once is allowed. There would have to be a coincidence of scales for
there to be a sizable branching fraction for the neutralino to each
of these, instead of one mode dominating. Distinguishing between these
sources of highly displaced jets is left to the next section, but
for now it should be noted that these types of searches have already
been considered in the literature for gravitinos~\cite{Meade:2010ji}
and RPV~\cite{Graham:2012th}, and these studies can be used as a guide for what
can be done with axinos. Besides removing the SM background, highly
displaced jets also allow the SUSY production channel and the signal
to be discussed independently. Regardless of how neutralinos are produced,
either in a simple two to two process, or at the ends of various long
cascade chains, the multi-jet signal is relatively unchanged, so long
as the displaced jets are what is triggered on. This means that optimistically,
the rate of the displaced jets signal can be taken as the inclusive
SUSY production rate for a given benchmark. There are however, a few
ways the production mechanism will effect the signal, even for highly
displaced jets. Exclusive neutralino pair production will produce
the most highly boosted jets, with longer and longer decay chains
reducing the amount of boost, though this is likely a small effect.
In addition to this distribution of boosts, the rest of the SM particles
produced in decay chains must be considered when determining the MET
of the whole event, and the MET resolution may vary between decay
chains. Another effect to consider is that the triggers for highly
displaced objects usually have isolation requirements, so that the
production channel for neutralinos must not produce calorimeter activity
in a region that points to the displaced decay.

While there are these advantages to considering highly displaced jets,
the drawback is that jet measurements may be difficult in the outer
parts of the detector. The degree to which detailed reconstruction
of jets in the outer detector is possible is beyond the scope of this
work, but at least it should be noted that in similar searches, such
as displaced decays to gravitinos \cite{Meade:2010ji}, the strategy
is to make use of triggers developed for hidden valley searches at
ATLAS \cite{Aad:2013txa}. The hope is that these same triggers could
be used for displaced decays to axinos. ATLAS has an advantage here
simply because of the detector geometry: a larger detector has a chance
to detect particles with longer decay lengths. A hidden valley can
produce displaced jets very similar to gravitinos or axinos, but the
hidden valley is not a particular model, or even frame work of models,
but rather a feature that can arise in various settings, so no attempt
is made in this work to make a direct comparison between a hidden
valley signal and other neutralino decays.

\begin{figure}
\centerline{\includegraphics[scale=0.2]{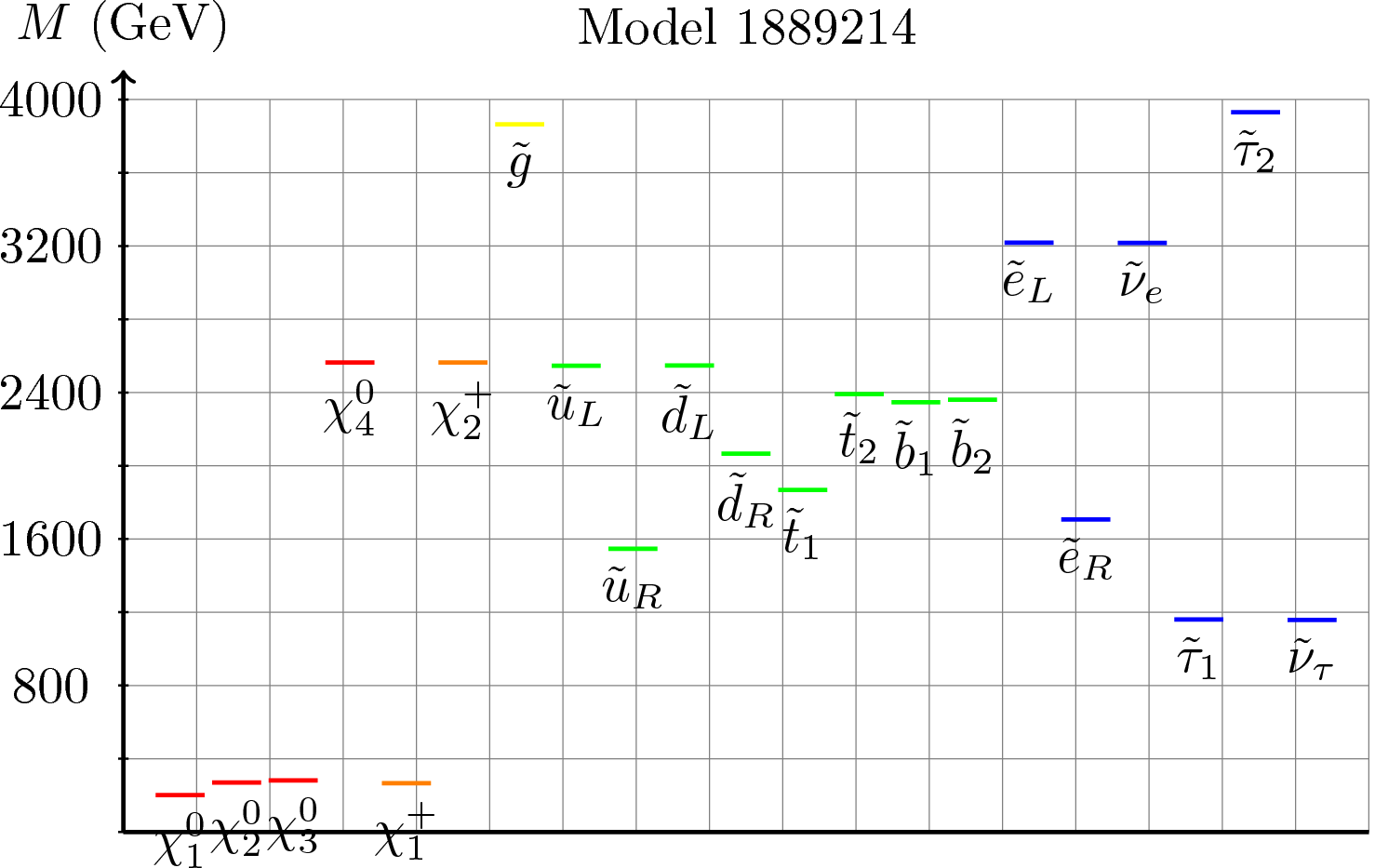}}
\caption{The ``lighter'' benchmark with an LOSP bino at $\approx200$~GeV,
taken from \cite{Cahill-Rowley:2013gca}. \label{fig:five}}
\end{figure}

\begin{figure}
\centerline{\includegraphics[scale=0.2]{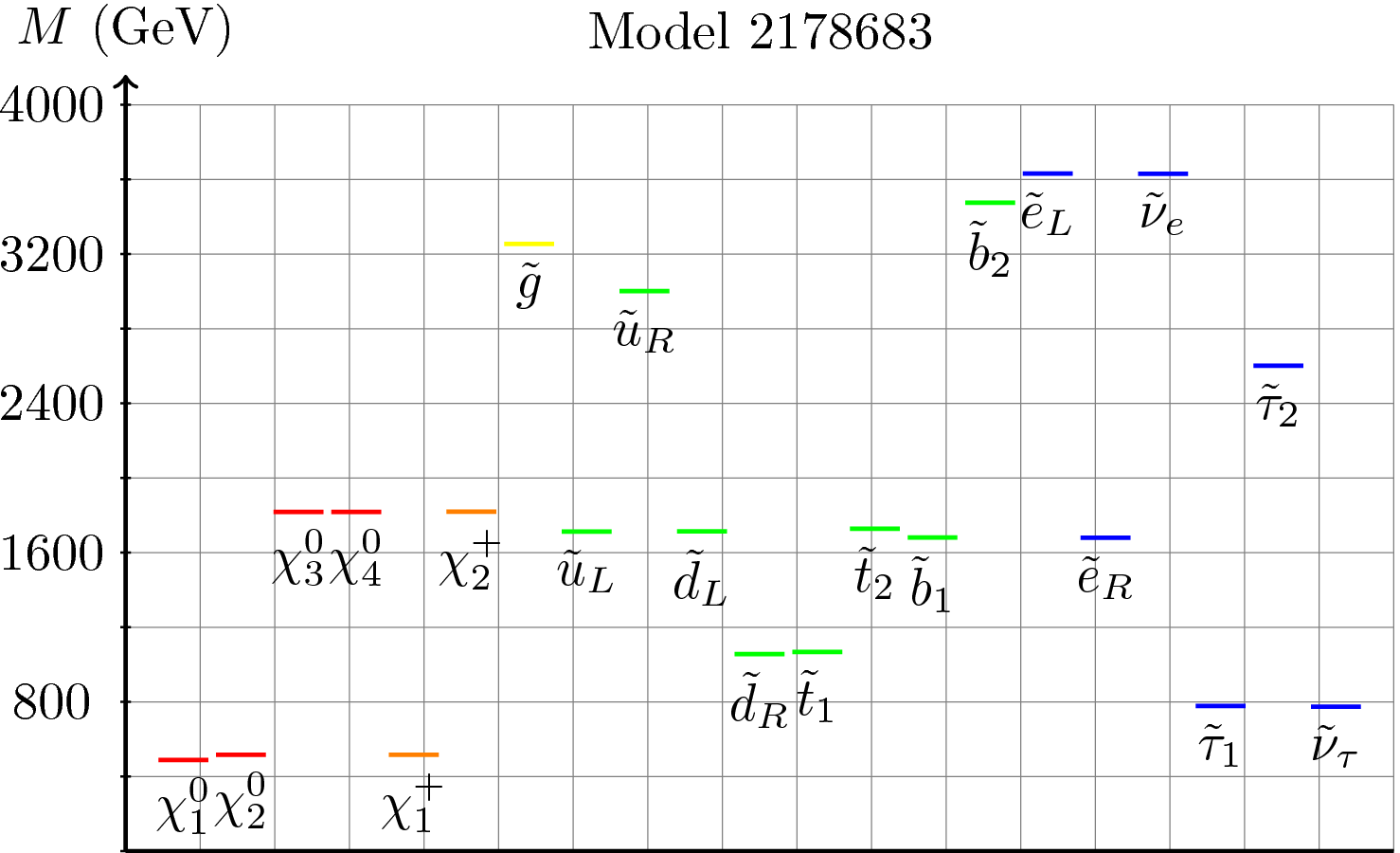}}
\caption{The ``heavier'' benchmark with an LOSP bino at $\approx500$~GeV,
taken from \cite{Cahill-Rowley:2013gca}.\label{fig:six}}
\end{figure}

With the expected signal identified as displaced jets and MET, the
SUSY spectra must be specified to obtain more quantitative results.
An appropriate benchmark SUSY model should meet a few criteria. Two
such benchmark models are chosen here, so that in Section~\ref{sec:results}
the effect of varying kinematics on the distributions can be explored.
Model 188924 and 2178683, both proposed as PMSSM benchmarks for Snowmass
2013 \cite{Cahill-Rowley:2013gca} are appropriate and appealing for
several reasons. The spectra of these models are given in Figures~\ref{fig:five}
and~\ref{fig:six}. The important difference between these two models
is that model 188924 has a lighter LOSP neutralino, a bino near 200
GeV, and here it will be referred to as the ``lighter'' benchmark,
while model 2178683 also has an LOSP bino, but a bit heavier, closer
to 500 GeV in mass and will be referred to as the ``heavier'' benchmark.
Both models have colored sparticle masses all between 1~TeV and 4~TeV.
These masses are the relevant model parameters to the topology in
Figure~\ref{fig:four}, along with the neutralino mixing and the
PQ scale, $f_{a}$.

The 19 parameter PMSSM makes no assumptions about the high scale theory
of supersymmetry and only specifies parameters at a low scale. While
it is more flexible then a SUSY model that specifies how SUSY is broken,
such as Minimal Super Gravity (MSUGRA) or GMSB, it can also contain
these models as a subset. Besides being agnostic to high scale physics,
this set of benchmarks was chosen by the authors as being testable
at a 14 TeV LHC and has been tested thoroughly so that it evades the
gauntlet of existing searches up to this point. Many SUSY models can
evade existing constraints, but this model does not implement any
special considerations to do so, it is simply the result of a scan
of the large PMSSM parameter space, and so can be thought of as a
``generic'' SUSY model that may be realized in the next run of the
LHC. All the models in this collection are stated to have possible
dark matter candidates, in that they do not over saturate the relic
abundance, but this point is moot in this context since the neutralino
LOSPs will all decay to axinos in the scenario here. In addition to
these features which are common to all of the PMSSM benchmarks described
in \cite{Cahill-Rowley:2013gca}, the benchmarks for this specific
scenario of neutralino decays to axinos requires a few more features.
The total neutralino width should be in a range such that the decay
is clearly displaced from the primary vertex, but still within the
ATLAS detector. The range considered appropriate for this is between
0.1~m and 10~m. The analysis is also easier if only one of the two
possible decays (two jet or three jet per leg) is clearly dominant,
so that there is no issue of double counting and matching with the
number of jets. For both the lighter and heavier benchmarks chosen
here with a gluino heavier than most of the squarks, the 3 jet channel
is suppressed by several orders of magnitude compared to the 2 jet
channel, so that the only coupling that needs to be considered is
${\cal L}_{\tilde{a}q\tilde{q}}$ of Eq.~\ref{eq:aqq} and the only
relevant topology is that shown in Figure~\ref{fig:four}. Also,
with neutralinos in this mass range, the branching fraction to heavy
quarks is greatly suppressed so that the neutralino branching fraction
to two light jets and an axino is very nearly one. There should also
be an adequate rate for a signal, which optimistically can be taken
as the inclusive SUSY rate. At the 14 TeV LHC the total inclusive
SUSY cross section for the lighter model benchmark is $\sigma_{SUSY}=5.4$~fb,
and for the heavier one is $\sigma_{SUSY}=23$~fb, as obtained at
leading-order with MadGraph/MadEvent \cite{Alwall:2014hca}. Several
of the benchmarks in this collection actually satisfied all of these
criteria, and the lighter benchmark model 188924 was only chosen because
it has the added appeal of relatively light sparticles, especially
with relatively light Higgsinos near 270 GeV, indicating that this
benchmark may have lower tuning. The heavier benchmark was simply
chosen because a heavier neutralino will have an impact on the kinematic
distributions used to distinguish between different neutralino decays
(axino/gravitino RPV) as will be shown in Section~\ref{sec:results}.

Several SUSY masses and parameters are not specified in these benchmarks.
The gravitino, for the purposes of this study, will be assumed to
be heavy enough that it does not effect the collider phenomenology.
Note that this does not have to be very heavy, as only very light
gravitinos are expected to not escape the detector. Gravitinos heavy
enough to decouple from collider physics can still have a large effect
on the cosmology. The only remaining masses are those from the axion
supermultiplet. The KSVZ axions mass is directly determined by the
scale $f_{a}$ so in the hadronic axion window these axions are still
very light, with a mass of approximately 10 eV. The axion is still
too weakly coupled to have an effect on collider studies, and since
it is R-parity even, there are no tricks to apply as in the case of
the axino. The scalar saxion's mass is model dependent, but it is
not expected to effect collider phenomenology, because like the axion,
it has even R-parity. Like the gravitino, the saxion can still greatly
affect the cosmology without changing collider studies. Finally, the
object of interest, the axino does not yet have a specified mass.
Theoretically the axino mass is highly model dependent and a large
range of values are explored in the literature, so it can be taken
as a free parameter here. As an LSP, lighter axinos are preferred
so they are not over-produced in the early universe \cite{Baer:2010gr},
but this will be somewhat alleviated if we assume a low reheat cosmology
as mentioned in the previous section. The signal of displaced jets
and MET is expected to be insensitive to the axino mass for relatively
light axinos. As the axino becomes heavy enough the width of the NLSP
will be affected, which will be explored in more detail in Section~\ref{sec:width}.

The signal for KSVZ axinos with a neutral NLSP at the LHC is very
predictive in that there are only two couplings to consider, each
providing one dominate topology. For this first study, the benchmarks
chosen have spectra that seems appropriate in that they are relatively
generic (the result of scans and not specifically ``engineered'')
and they are expected to give decay lengths in an appropriate range.
Beyond these particular benchmarks, the mixing of the NLSP neutralino,
the amount of compression of the spectra, and the mass of the axino
may affect the signal to some degree and it is worth testing. Though
there are no SM backgrounds to compete with very displaced jets, there
are other possible decays for the neutralino, including decays to
gravitinos and decays via RPV. Now that these qualitative aspects
of the signal have been summarized, in the next section results are
presented for simulated events, for decays to axinos and the alternatives.
The impact of the possible effects described above are explored and
the degree to which these neutralino decays can be distinguished is
tested.

\section{Results and Discussion}

\label{sec:results}

To investigate how predictive the axino multi-jet and MET signal is,
we simulated events for the LHC at 14 TeV. The primary tool used to
generate Monte Carlo events was MadGraph/MadEvent
\cite{Alwall:2014hca}. The parton distribution function (PDF) set used
with MadEvent was CTEQ6L1~\cite{Pumplin:2002vw}. The renormalization and factorization
scales were allowed to run and were determined by MadEvent's default
settings with the scale for decay events set to the mass of the
decaying parent particle. We added the axino field and its couplings
to the MSSM using FeynRules
\cite{Alloul:2013bka,Degrande:2011ua,Duhr:2011se}. The FeynRules
implementation of the axino was validated for ${\cal
  L}_{\tilde{a}\tilde{g}g}$ of Eq.~\ref{eq:agg} by comparing the tree
level decay of a heavy axino to the analytical result, and for ${\cal
  L}_{\tilde{a}q\tilde{q}}$ of Eq.~\ref{eq:aqq} by comparing the
squark decay width to the results in \cite{Covi:2002vw}. We also
calculated the neutralino decay width for $\tilde{\chi}^{(}0)_{j}\to
q_{i}\bar{q}_{i}\tilde{a}$ analytically and confirmed the result of
~\cite{Barnett:1987kn} with the appropriate adjustments (see
Eq.~\ref{eq:width}), and used this analytic form of the decay width to
verify the results obtained with MadGraph/MadEvent.

Existing model files in the FeynRules data base were used when
generating comparison events for the cases with gravitinos
\cite{Christensen:2013aua} and RPV couplings \cite{Fuks:2012im}. Mass
spectra were generated using SoftSusy \cite{Allanach:2001kg} and
checked with SuSpect \cite{Djouadi:2002ze}.  Jet clustering was done
with FastJet \cite{Cacciari:2011ma} using $kT$ jets with $D=0.4$, and
parton showers were generated by Pythia \cite{Sjostrand:2006za}. The
analysis is done in Mathematica with the Chameleon package
\cite{Thaler:2006:Online} as a base, but with plenty of modifications
and extensions.  Examples of Mathematica notebooks for event analysis
with the Chameleon package including these modifications can be found
at \cite{Redino:2015:Online}. Events are generated at tree level, but
the vertex correction of Fig.~\ref{fig:three} is captured in the
effective coupling in ${\cal L}_{\tilde{a}q\tilde{q}}$ of
Eq.~\ref{eq:aqq}.

The only tool required for this study which is less common was evchain
\cite{Kim:2013ivd} which acts as a ``MadGraph manager'' to combine
separate subprocess runs, and is especially useful for decay chains
which are difficult for MadGraph to manage alone. In this scenario
with axinos, MadGraph has difficulty because of the extremely narrow
decay width of the neutralino. As described in the previous section,
there is effectively only one topology by which the neutralino can
decay, i.e. to two jets and the axino. While the branching fraction
to two jets and the axino is very close to one, the width is still
extremely small because of both the suppression from the presence
of $f_{a}$ in the denominator and because of the heavy off-shell
squark required for the decay. MadGraph can generate the decays of
the neutralino just fine, but to include these decays in a larger
event is problematic.

Looking at just the neutralino decay alone, the decay width is
calculated from which the expected decay length $c\tau$ in the
detector is determined, so the assumption that there are plenty of
highly displaced jets can be tested. For the lighter benchmark of
Figure~\ref{fig:five} with a light axino, the width of the lightest
neutralino varies between
$\Gamma_{\tilde\chi^{0}_1}=7.3\times10^{-16}$~GeV and
$\Gamma_{\tilde\chi^{0}_1}=1.7\times10^{-17}$~GeV over the window
$3\times10^{5}{\rm GeV}<f_{a}<3\times10^{6}{\rm GeV}$.  This
corresponds to a mean decay length range between roughly
$c\tau=0.26$~m and $c\tau=11.6$~m. For the heavier model
Figure~\ref{fig:six} over the same range in $f_{a}$ the neutralino
width spans the range
$\Gamma_{\tilde\chi^{0}_1}=1.7\times10^{-13}$~GeV and
$\Gamma_{\tilde\chi^{0}_1}=1.2\times10^{-15}$~GeV or a length range of
$c\tau=0.0012$~m and $c\tau=0.16$~m. This range is a very appropriate
size for the ATLAS detector, allowing for a sizable number of events
that are displaced enough to realize the advantages described in the
previous section: negligible SM backgrounds, and the ability to
separate particles from production and particles from neutralino
decay. This width is also insensitive, i.e. within statistical errors,
to the axino mass in the range $0\le m_{\tilde{a}}<10$ GeV. The hope
is that the axino signal would be trigger-able at this depth in the
ATLAS detector using the hidden valley triggers discussed in
\cite{Aad:2013txa}. No serious attempt is made here at determining the
efficiency of such triggers for this model, as adjusting for instance,
detector simulation tools for displaced jets is non-trivial work, and
not readily available in off-the-shelf tools, but Meade {\it et al.} do make
an estimate of the efficiency of some of these triggers for highly
displaced jets in \cite{Meade:2010ji}.

For the remainder of events analyzed the axino is assumed to be very
light (effectively massless) and $f_{a}=10^{6}$~GeV, corresponding to
a neutralino decay width of
$\Gamma_{\tilde\chi^{0}_1}=1.1\times10^{-16}$~GeV for the lighter
benchmark and $\Gamma_{\tilde \chi^{0}_1}=7.5\times10^{-15}$~GeV for
the heavier benchmark.

If the trigger can actually be agnostic to the production mechanism,
then all SUSY channels can contribute to the signal cross section, and
for the benchmark this gives an inclusive SUSY rate of
$\sigma_{SUSY}\sim5$~fb for the lighter benchmark and
$\sigma_{SUSY}\sim23$~fb for the heavier one. In the much more
pessimistic case where we only attempt to look for events with
neutralino decays only, i.e. neutralino pair production, then the rate
is only $\sigma_{\tilde\chi^{0}_1\tilde\chi^{0}_1}\sim30$~ab for the
light benchmark and
$\sigma_{\tilde\chi^{0}_1\tilde\chi^{0}_1}\sim14$~ab for the heavy
one, possibly providing just a few events with ${\cal L}=300\,{\rm
  fb}^{-1}$, if they survive the efficiency of the triggers (note that
the HL-LHC is designed to reach ${\cal L}=3\,{\rm ab}^{-1}$).

Looking at simulated events for the decay alone is still useful for
studying the shapes of the kinematic distributions. Even though the
neutralino decays actually will occur after some production process
with boosted momentum and convolution with PDFs, the distributions
from decay-only events are still physical in that they show the
relevant observables in the neutralino rest frame. These rest-frame
distributions can provide interesting hints as to how the lab-frame
distributions may be distinguished between different neutralino decays
(axino/gravitino/RPV). More optimistically, these rest-frame
distributions may be directly accessible, if the neutralino momentum
in the lab frame can be reconstructed, then the appropriate boost on
the lab-frame observables can be made. Such a boosting is not a simple
task for partially invisible decays, and no explicit algorithm is
provided here, but similar reconstruction for partially invisible
decays has been done for instance in the context of top decays
\cite{Guillian:1999jh}.  Cleaner distributions (without PDF
convolution) would also be accessible at a lepton collider, since this
signal channel is indifferent to the SUSY production mechanism.

When looking at the full event, with SUSY production and the full
decay, with such a small decay width, MadEvent fails to sample an
appropriate phase space and the results of the Monte Carlo integration
are unreliable. This can be illustrated as follows: in the narrow-width
approximation, 
\begin{equation}
d\sigma_{tot}=d\sigma_{prod}\frac{\Gamma_{decay1}}{\Gamma_{total}}\frac{\Gamma_{decay2}}{\Gamma_{total}}\label{eq:nw}
\end{equation}
where here BR$=\Gamma_{decay1/decay2}/\Gamma_{total}\sim1$ for both
decays and a very narrow neutralino decay width $\Gamma_{total}/m_{\tilde\chi^{0}_1}\ll1$,
the cross section of neutralino pairs should be the same, regardless
of whether or not their decays are included, that is $d\sigma_{tot}=d\sigma_{prod}$,
which is not found with MadEvent when including the neutralino decays
in this model. The way evchain circumvents this limitation is in a
way by implementing the narrow-width approximation ``by hand''.
The production process for neutralinos is done in one run of MadGraph
(either by direct pair production or via any SUSY cascade) and the
decay of the neutralinos is done in another, separate run. The resulting
LHE event files from these two separate runs are combined by evchain
(with the appropriate Lorentz boosts being made), and the cross section calculated
from production events is scaled by the branching fraction to the
decay events, as per the narrow width approximation of Eq.~\ref{eq:nw}.
In the case of the axino LSP, no scaling is necessary since the branching
is effectively one, but when similar events with gravitinos and RPV
decays are generated for comparison, the appropriate scaling has to
be applied.

A minimal set of loose default generation cuts are implemented in
MadEvent (with any other cuts done during the analysis). These cuts
include a minimum jet $p_T$ of 1~GeV, a minimum invariant mass between
any pair of jets of 1 GeV and a delta R between jets of 0.1 This set
of cuts is the same for all three models (axino/gravitino/RPV).  It
should be noted that these cuts themselves, are ``boosted'' by evchain
as well, for example a small minimum $p_{T}$ requirement on a jet will
actually cut events at a higher $p_{T}$ after evchain boosts the
events. This effect should be negligible however, as in all the events
generated the $p_{T}$ of jets is rather large (also good news for
triggering) because of the mass of the neutralino NLSP.

The comparison models are intended to be as similar as possible to
the benchmark cases for axinos so that distinguishing between events
here can be thought of as a ``worst case'' scenario, where distinguishing
models is the most difficult. This also means that the axino is taken
to be very light, like the gravitinos or neutrinos (from RPV) that
appear in the alternative neutralino decays. A heavier axino will
provide more handles for distinguishing neutralino decays via the
kinematic distributions. Of course the ability to distinguish between
distributions is dependent on the ability of the detector to measure
such features when the jets are highly displaced, and again, no detailed
detector simulation is attempted in this work. Aside from the axino
mass and the gravitino mass, the rest of the SUSY spectra is identical
between the models, so the production rates above are the same for
all three alternatives (axino/gravitino/RPV). In addition to the similar
spectra, the width of the lightest neutralino to 2 jets and MET is
made to be very close to the axino benchmarks so that the similar
signal would appear in the same region of the detector (though not
necessarily at the same rate, since the total width does not have
to be the same as the axino benchmark). It is reasonable to assume
that other comparison models could produce better ``imposter'' signals,
by having a higher rate (from a different SUSY spectra) and a different
decay length, but with a comparable number of events in the same part
of the detector. Comparison of the axino benchmark signal to these
two comparison models follows, with gravitinos first, and then with
the RPV signal.

\begin{figure}
\centerline{\includegraphics[width=4.5in]{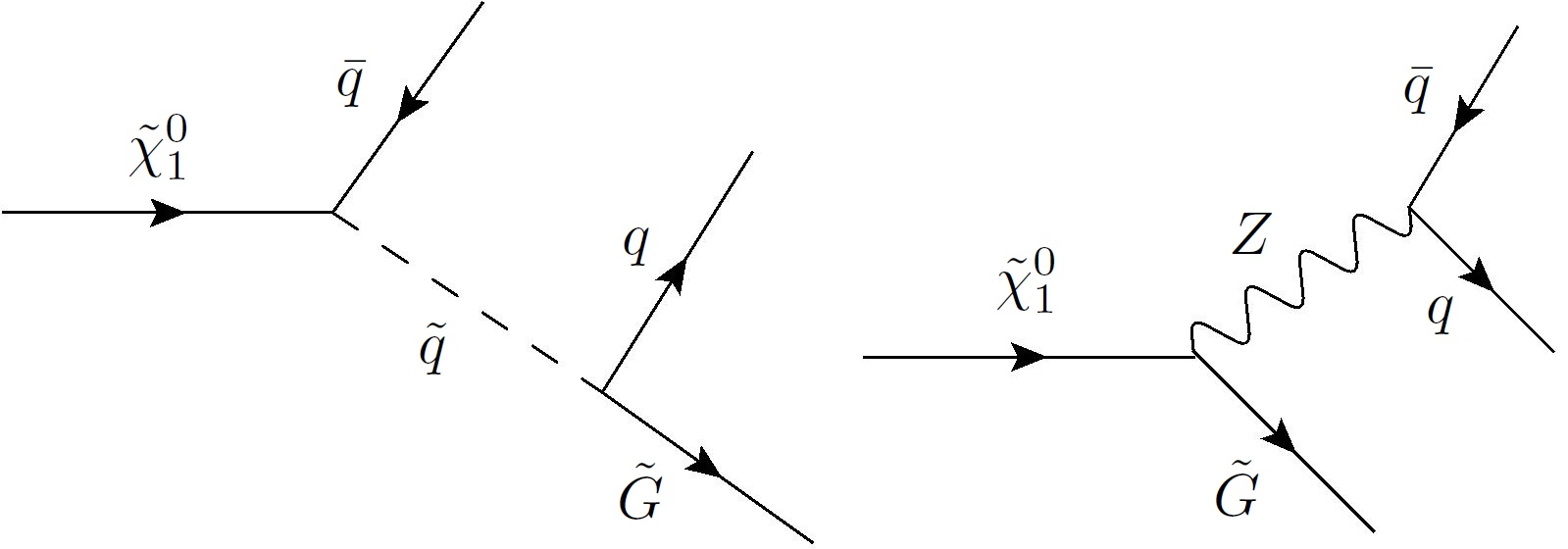}} \caption{Two diagrams that contribute to neutralinos decaying to two jets and
a gravitino. \label{fig:seven}}
\end{figure}

\begin{figure}
\centerline{\includegraphics[width=5.5in]{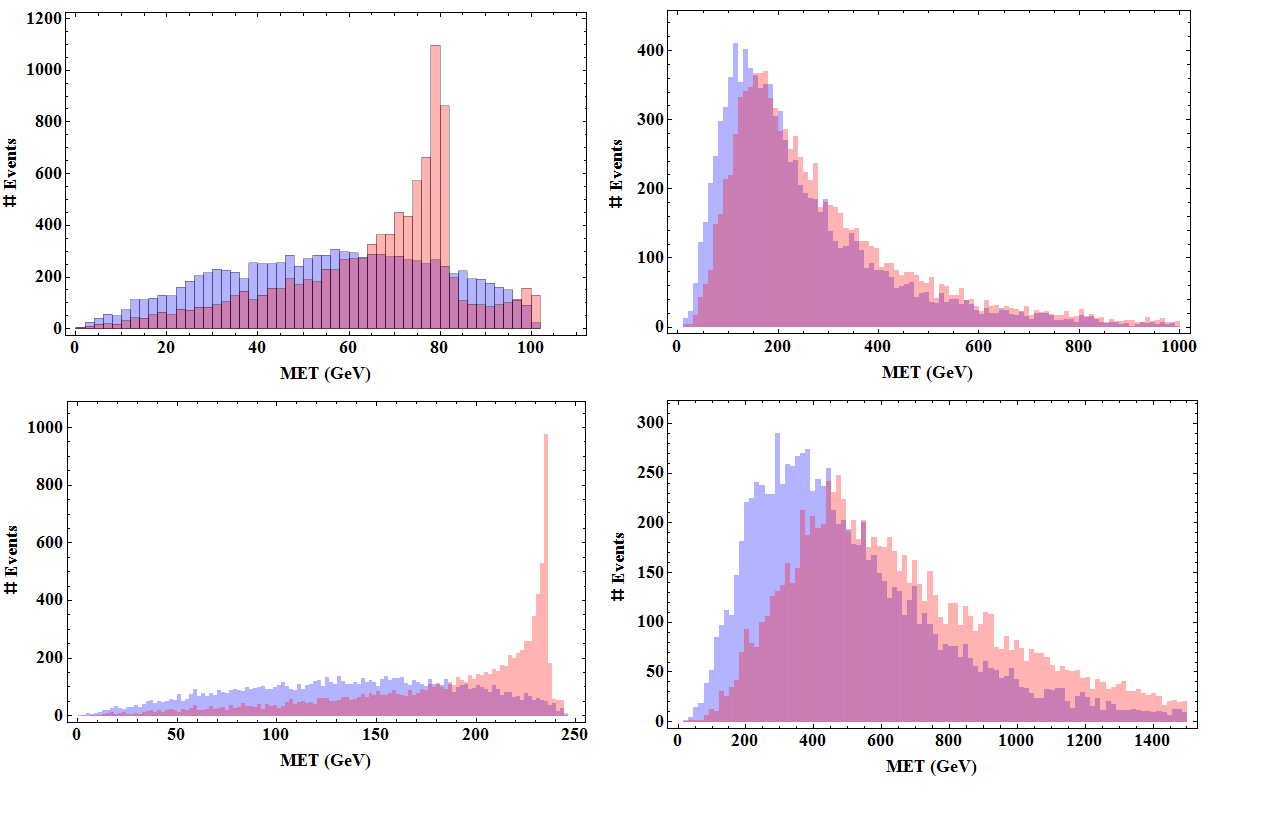}}
\caption{Distributions of the MET from neutralino decays to axinos (blue) and
gravitinos (red).  Events are simulated with minimal generation cuts only, and at parton level (no showering/clustering). The left plots  consider the neutralino decay alone, while the right plots are in the lab frame
of the whole event at 14 TeV at the LHC, i.e. when including both production and decay via evchain. The upper
  plots are for the lighter benchmark, and the lower plots for the
  heavier benchmark. \label{fig:eight}}
\end{figure}

\begin{figure}
\centerline{\includegraphics[width=5.5in]{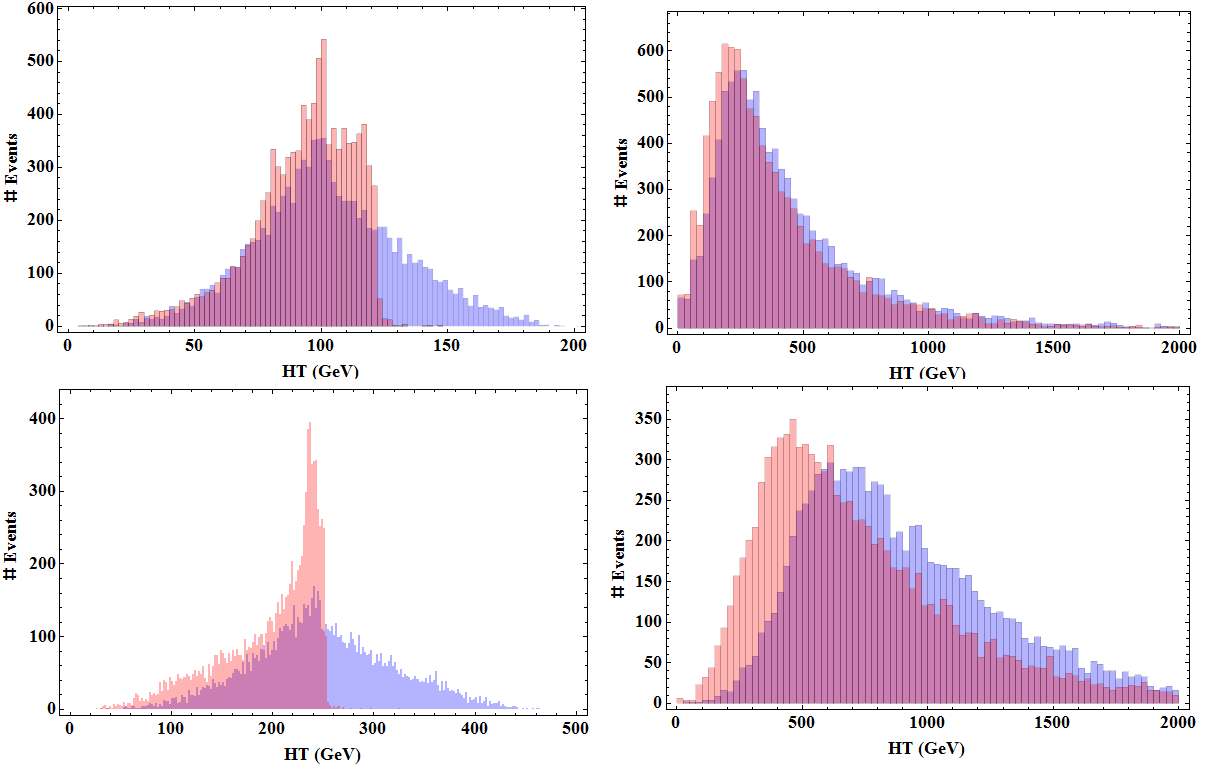}}
\caption{Distributions of the scalar sum of jet transverse momenta (the $H_T$) from
  neutralino decays to axinos (blue) and gravitinos (red). Events are simulated with minimal generation cuts only, and at parton level (no showering/clustering). The left plots  consider the neutralino decay alone, while the right plots are in the lab frame
of the whole event at 14 TeV at the LHC, i.e. when including both production and decay via evchain. The upper
  plots are for the lighter benchmark, and the lower plots for the
  heavier benchmark. \label{fig:nine}}
\end{figure}

\begin{figure}
\centerline{\includegraphics[width=5.5in]{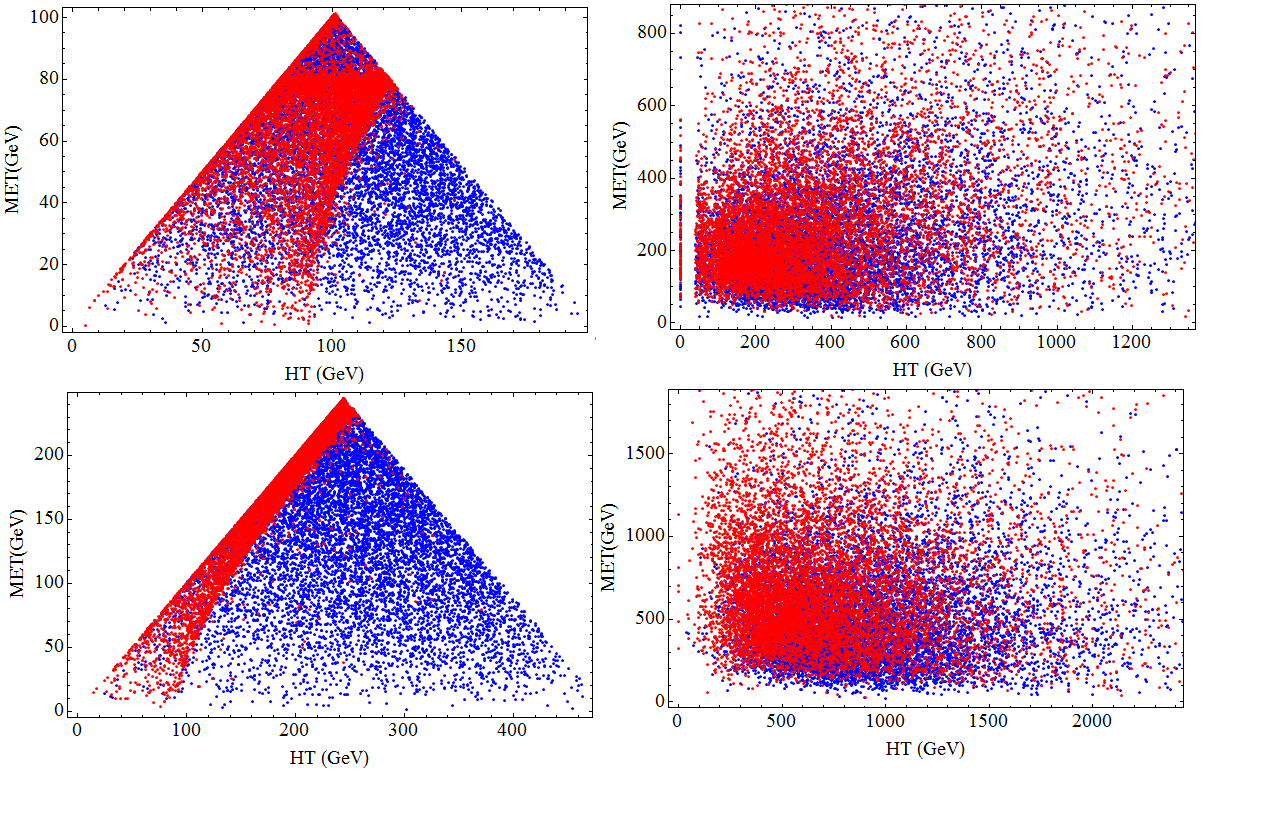}}
\caption{$H_T$ versus MET from neutralino decays to axinos (blue) and
  gravitinos (red).  Events are simulated with minimal generation cuts only, and at parton level (no showering/clustering). The left plots  consider the neutralino decay alone, while the right plots are in the lab frame
of the whole event at 14 TeV at the LHC, i.e. when including both production and decay via evchain. The upper
  plots are for the lighter benchmark, and the lower plots for the
  heavier benchmark.  \label{fig:ten}}
\end{figure}

In the case of gravitinos, to have a neutralino decay length for the 2
jet and MET signal similar to the axino model, the gravitino mass is
taken to be $500$~eV for the lighter benchmark and $750$~eV for the
heavier one. Unlike the axino, the gravitino has numerous couplings,
and is not restricted to the 2 jet+MET channel. One obvious
consequence of this is that gravitinos can be distinguished from
axinos simply by looking for these other decay channels, e.g. anything
with leptons or photons that is highly displaced. There is plenty of
literature describing how to search for gravitinos with leptons or
photons, see, e.~g., ~\cite{Aad:2014gfa}, but this is not enough to
rule out the possibility that some of the displaced multi-jet signal
could be coming from axinos.  While it would require a coincidence of
parameters, neutralino decays to gravitinos and axinos could co-exist
in a model, so for the sake of being thorough, the 2 jet+MET signals
can be compared in hopes of distinguishing them based on the shapes of
kinematic distributions alone. By the same reasoning, the presence of
alternative decay channels for the gravitino means that the branching
fraction will be less than one, and so for identical SUSY spectra the
gravitino model will have the multi-jet+MET signal occurring at a
smaller rate. Unlike the axino case, there are several topologies to
produce two jets and a gravitino (see Figure~\ref{fig:seven}). While
the same topology exists as in the axino case (Figure~\ref{fig:seven}
left), it is not dominate here, as there are also topologies without
off-shell colored sparticles that contribute with much more strength
(Figure~\ref{fig:seven} right).  Diagrams like the right one in
Figure~\ref{fig:seven} give two hints as to how the scenarios can be
distinguished. Since the dominant topology has the gravitino radiated
at the beginning of the decay chain, rather than the end, we should
expect the MET to recoil differently between the axino and gravitino
models, with the gravitino MET being harder. Conversely, the visible
jets should be harder for the axino case, and softer for the decays to
gravitinos. These kinematic differences are subtle when looking at
neutralino decays for this particular benchmark, and when these decays
are simulated for a SUSY cascade the effect is washed out by the
boosts and smearing by the PDFs. For a model with a much heavier
neutralino this difference is more noticeable. Results comparing the
MET for axino events versus gravitino events are shown in
Figure~\ref{fig:eight} and results comparing the total jet $p_{T}$
(the $H_{T}$) are shown in Figure~\ref{fig:nine}. Jet $p_{T}$ and MET
being relatively larger or smaller between decays with axinos or
gravitinos is not a very useful handle by itself. If only one type of
decay is actually measured, it begs the questions: more or less
$p_{T}$ compared to what? The difference in weighting between visible
and invisible $p_{T}$ can be seen by plotting $H_{T}$ against the MET,
as shown in Figure~\ref{fig:ten}.  In all of these kinematic plots the
effect is more noticeable for larger neutralino masses, but it is
still smeared away in the full event, i.~e. when including both
production and neutralino decays.

\begin{figure}
\centerline{\includegraphics[width=5.5in]{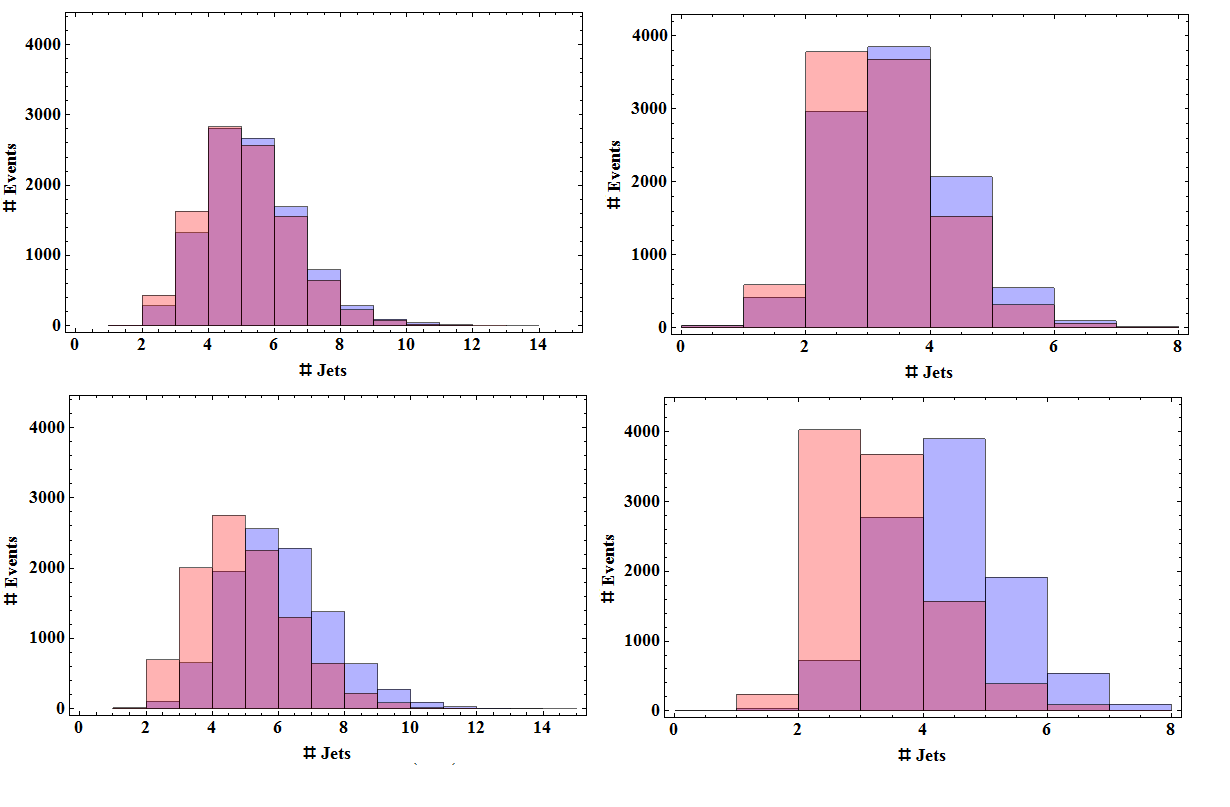}}
\caption{The number of jets from neutralino decays to axinos (blue) and gravitinos
(red). All plots are for the full event (production and decay via evchain) at 14 TeV with showering done by PYTHIA and jet clustering from FastJet using $k_T$ jets with $D=0.4$. The left plots are from events generated with loose generation cuts and the right plots
are obtained after applying more restrictive cuts, $p_T>40$~GeV and $|\eta|<2.5$. The
upper plots are for the lighter benchmark, and the lower plots for
the heavier benchmark. \label{fig:eleven}}
\end{figure}

Perhaps a simpler way to distinguish these two models is to just count
the displaced jets. It is reasonable to think that after parton showering
with Pythia and clustering we could expect more jets from the axino case as these
jets are expected to be harder (based on the kinematic distributions
above) and are more likely to radiate, and this is reflected in the
generated events as shown in Figure~\ref{fig:eleven}. Applying stronger
jet $p_{T}$ cuts to satisfy triggers~\cite{Aad:2013txa} ($p_{T}>40{\rm GeV})$
will remove the softer jets from both samples, in addition to a pseudo-rapidity
cut ($|\eta|<2.5)$. This does not just reduce the total number of
jets after showering, but it also shifts events between different
bins for numbers of jets.

\begin{figure}
\centerline{\includegraphics[width=5.5in]{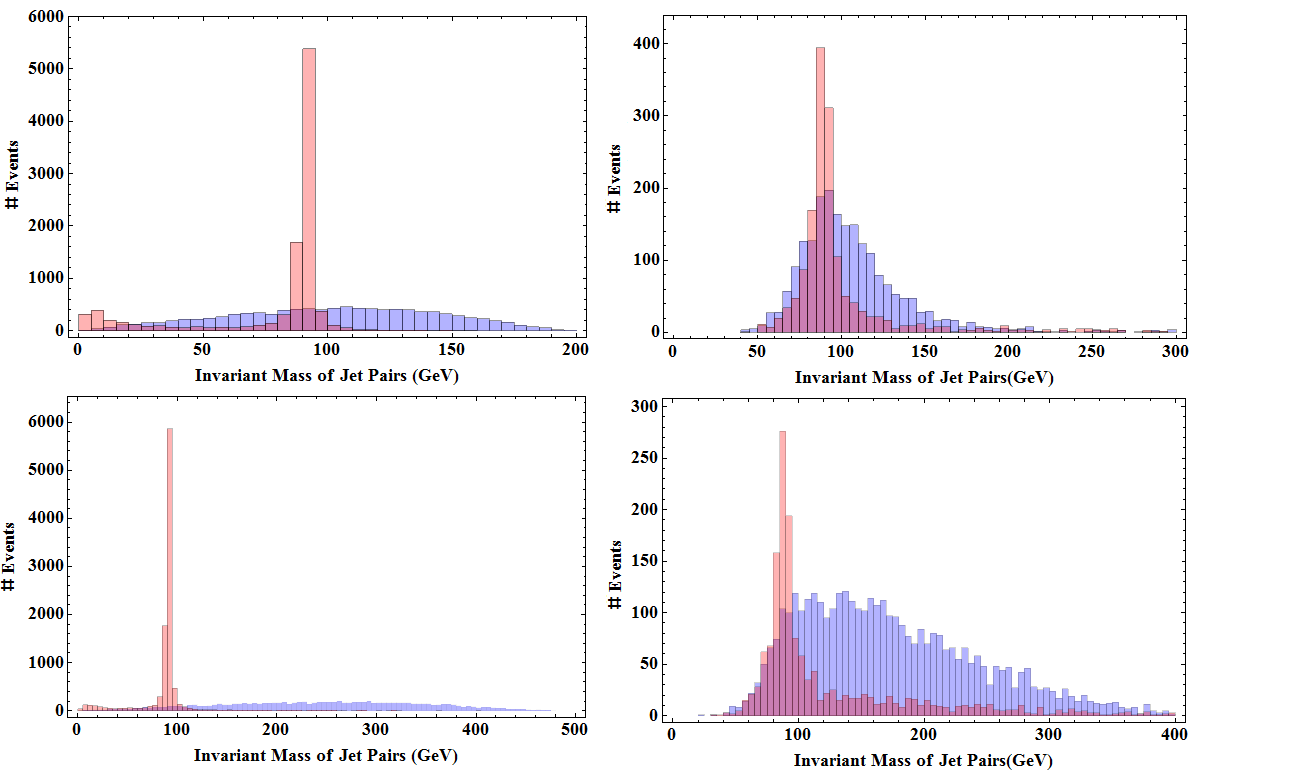}}
\caption{The invariant mass of jet pairs for neutralino decays to
  axinos (blue) and gravitinos (red). The left plots are for a single
  decaying neutralino in its rest frame, while the right is in the lab
  frame for the whole event (in a sample selected to have exactly four
  jets) at 14 TeV with showering done by PYTHIA and jet clustering
  from FastJet using $k_T$ jets with $D=0.4$. The upper plots are for the
  lighter benchmark, and the lower plots for the heavier
  benchmark. \label{fig:twelve}}
\end{figure}

Even though a photon or a Higgs boson could take its place in diagrams
similar to the left diagram in Figure~\ref{fig:seven}, the presence of the $s$-channel
$Z$ has a significant effect on the distributions and the $Z$ resonance
can be reconstructed. For event samples from just the decays, the
$Z$ resonance simply comes from the invariant mass of both jets (Figure~\ref{fig:twelve}).
When looking at the full event with showering there is the question
of which jet combination to take. Including the full combinatoric
background (all combinations of jets), the $Z$ resonance is buried,
but since the particular resonance is known in this case, it is easy
enough to just take those combinations of jets which are closest to
the known $Z$ mass. This method has the draw back that it can create
an artificial bump in the jet invariant mass distribution. This artificial
bump is much more pronounced when the parent neutralino is lighter,
so again, like the other methods of discrimination, it is more difficult
for lighter neutralinos. Reconstructing the $Z$ resonance is a much
more powerful way to distinguish the gravitino and axino cases, as
unlike the kinematic distributions discussed earlier it is invariant
to boosts and somewhat insensitive to the sparticle masses. A veto
on the invariant mass of jet pairs in separate halves of the detector
(pairs from separate parent neutralinos) can cut the majority of gravitino
events and of the methods described here such a veto is considered
the best way to determine if neutralino decays contain events with
axinos, gravitinos, or both.

\begin{figure}
\label{fig:thirteen} \centerline{\includegraphics[scale=0.3]{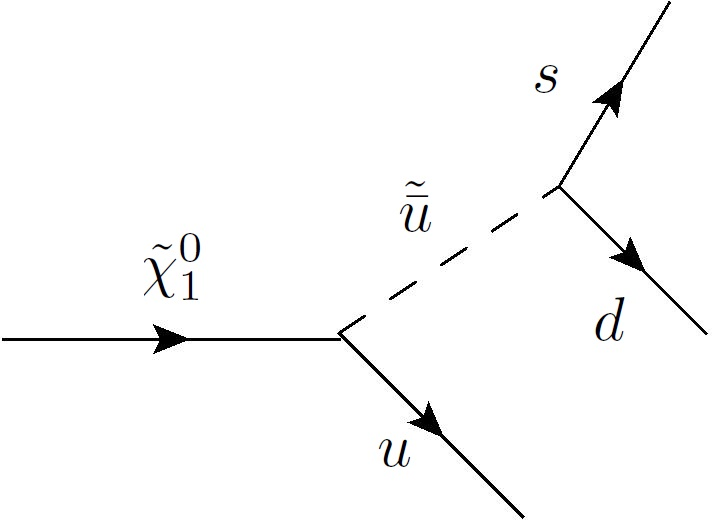}}
\caption{RPV topology with UDD coupling.}
\end{figure}

Like gravitinos, RPV scenarios can also produce a signal of displaced
jets and missing energy, and RPV decays can co-exist in a model with
decays to axinos, so a comparison of these similar signals is warranted.
There are many possible signatures of RPV as there are several possible
couplings, coming from both the super potential and also from soft
SUSY breaking terms. The couplings from the R-parity violating super
potential are given by~\cite{Barbier:2004ez}
\[
W_{RPV}=\mu_{i}H_{u}L_{i}+\frac{1}{2}\lambda_{ijk}L_{L}^{i}\centerdot L_{L}^{j}E_{R}^{k}+\lambda'_{ijk}L_{L}^{i}\centerdot Q_{L}^{jl}D_{Rl}^{k}+\frac{1}{2}\lambda''_{ijk}\epsilon^{lmn}U_{Rl}^{i}D_{Rm}^{j}D_{Rn}^{k}\;,
\]
where $i,j,k$ are flavor indices and $l,m,n$ are color indices.
The first three terms all violate lepton number, while the last term
violates baryon number. While all these couplings are possible, they
are constrained by the non-observation of certain processes. To avoid
running into bounds from unobserved processes, such as proton decay,
the constraint is not on the size of these couplings directly, but
rather of their products (for example proton decay requires B and
L to be violated). Because of this it is not unreasonable to assume
that there could be just one dominate RPV coupling, that is itself
relatively small. The UDD coupling can produce three displaced jets
from neutralino decay (Figure~\ref{fig:thirteen}), which may look
like the axino signal after showering/clustering but any missing energy
would have to come from detector/trigger inefficiencies, or jet mis-measurement.
It is difficult to estimate how much ``fake MET'' there could be
in such events without performing a detailed detector simulation,
but it is expected that such events would be distinguishable from
axino events for any model similar to the benchmarks, because the
NLSP neutralinos are massive enough and the axino will carry away
a large portion of this energy, as shown in the MET plots in Figure~\ref{fig:eight}. Also, due to the absence of true MET the jets themselves will be
harder than in the axino case.

In the case where the dominant RPV coupling is of the LQD type (with
$\lambda'\approx7.5\times10^{-6}$ for the lighter benchmark and
$\lambda'\approx2.5\times10^{-6}$ for the heavier benchmark), the two
jet signal can look very much like the axino case when the neutralino
decays to two jets and a neutrino.  As in the gravitino case, this RPV
coupling also allows for channels with photons and charged leptons in
the final state. Again, the presence of other channels will mean the
rate of the 2 jet signal is less than in the axino case, but a
difference of rates is not helpful without a priori knowledge of the
SUSY spectra to calculate these rates . Discovery of displaced photon
or charged channels would imply there are neutralino decays not
involving the axino, but again, as was mentioned with the gravitino
case, this does not exclude the possibility that both decays exist in
the same model. While it was stated that it would require a
coincidence of parameters to have competitive neutralino decays to
gravitinos and axinos in the same model, it would be less surprising
in the case of RPV with trilinear couplings, such as the UDD or LQD
ones explored here. Affleck-Dine baryogenesis (ADB) with RPV couplings
\cite{Higaki:2014eda} is a scenario that is attractively compatible
with a cosmology with LSP axinos in the hadronic axion window. This is
because with such a low Peccei-Quinn scale, $f_{a},$ the scenario
likely requires a very low reheating temperature, (which ADB can
accommodate, unlike thermal Leptogenesis). Also, when ADB involves RPV
couplings, there must be another source of dark matter instead of the
lightest neutralino, which axions/axinos can accommodate.  The
coincidence of scales required to make both RPV and axino decays
competitive comes from two independent sources. For the axinos, the
window of lower $f_{a}$ is set by the constraints mentioned in
Section~\ref{sec:background}, and for ADB with RPV to be successful it
requires a trilinear RPV coupling with
$\lambda\approx10^{-7}$\cite{Higaki:2014eda}, coincidentally in the
same range to give similar width as the axino decays. Though it is a
distinct possibility that these channels co-exist in the same model,
the ``coincidence'' should not be overstated, as depending on the
value of $f_{a}$ the width to axinos actual varies over a couple
orders of magnitude, and with RPV, ADB can be accommodated with
$10^{-9}<\lambda<10^{-6}$, so while these correspond to the same range
of widths for decays, it is also possible that one process dominates
and the other will have a negligible rate.

\begin{figure}
\centerline{\includegraphics[scale=0.3]{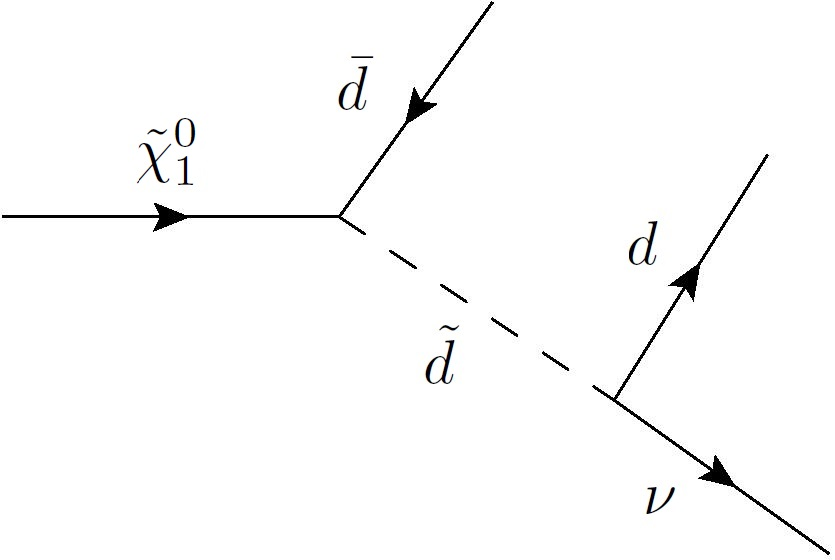}}
\caption{RPV topologies with LQD coupling. \label{fig:fourteen}}
\end{figure}

\begin{figure}
\centerline{\includegraphics[width=5.5in]{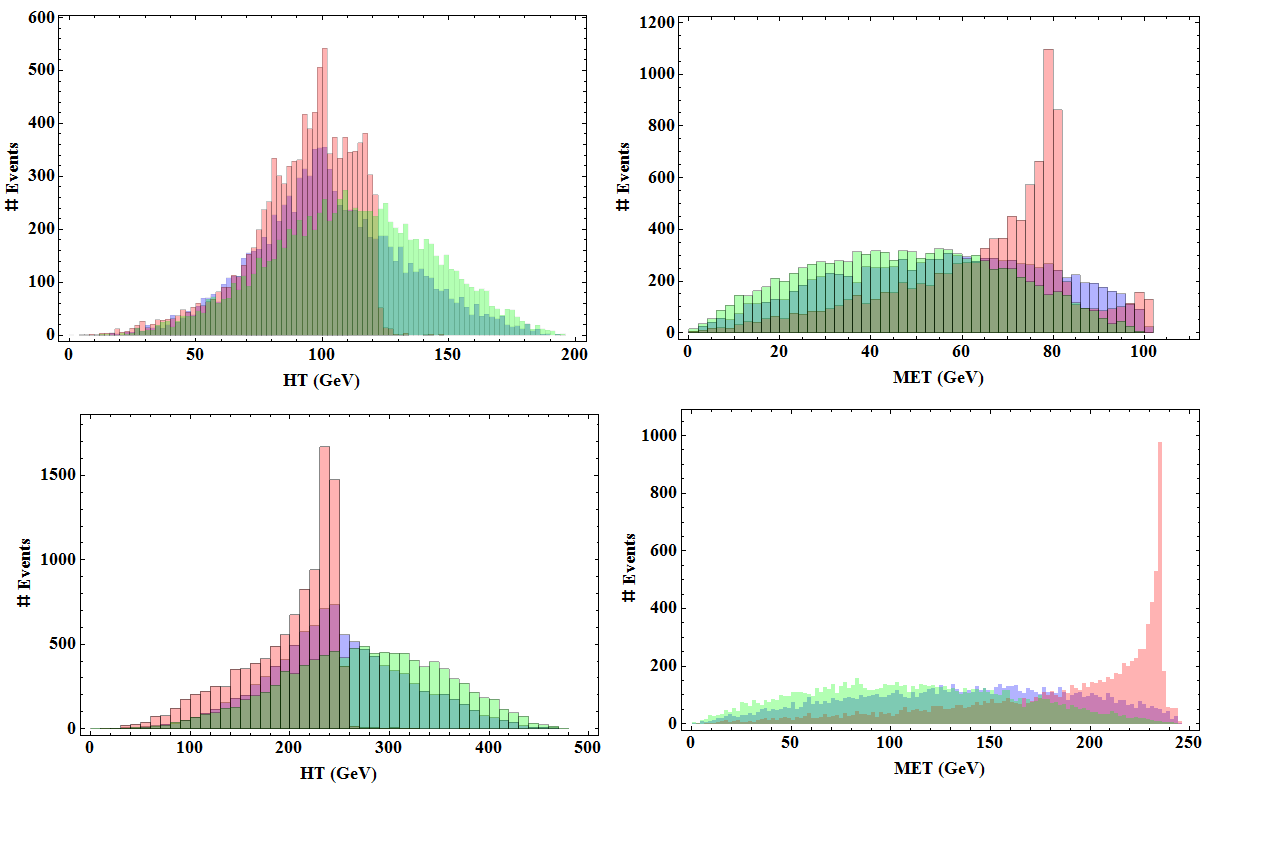}}
\caption{Summary of kinematic distributions for all three scenarios,
  axino (blue), gravitino (red) and RPV with an LQD type coupling
  (green).  Events are simulated for the decay only with minimal
  generation cuts only, and at parton level (no showering/clustering).
  The top row is the lighter benchmark and the bottom row is the
  heavier benchmark. \label{fig:fifteen}}
\end{figure}

Distinguishing LQD RPV from axino signals by the jet distributions
alone is much more difficult than the case with gravitinos as the
topologies contributing to the signal are now identical (Figure~\ref{fig:fourteen}).
There is no massive resonance to distinguish the models as in the
case of gravitinos, and the MET and various jet variables are also
very similar between this case and the axino case. Some of the distributions
explored earlier are shown again in Figure~\ref{fig:fifteen}, but
now with LQD RPV distributions as well. The only one that could potentially
be used as a tool to distinguish axino and RPV signals is the $H_{T}$,
which shows a peak at half the parent neutralino's mass for the axino
case, but not for the RPV case. This is only useful if the neutralino
mass is known or at least constrained (perhaps from analysis of the
prompt event separately) and the distribution is only useful in the
neutralino rest frame, so its momentum must also be determined. If
sufficiently long lived RPV decays are discovered at the LHC, it may
be very difficult to rule out the possibility of a light axino contributing
to some of that signal.

When considered together, there are a few handles to distinguish between
gravitino and axino scenarios, and with varying success, some RPV
scenarios, but it should also be asked how strongly any of these results
depend on the choice of benchmark. Most of the distinguishing features
above come about simply as a consequence of the topology, and should
not be very sensitive to many aspects of the benchmark. The parameter
that is expected to have the greatest effect on the shapes of these
distributions is the mass of the lightest neutralino, which has been
demonstrated in our results for the kinematic distributions. The choice
of neutralino mixing, in this case a mostly Bino NLSP, does not have
a large effect on any of the axino distributions shown here, and the
largest effect this has on the gravitino signal is to change the branching
ratio to jets, but the rates are already very different from the axino
case.

The more important effect from varying the SUSY mass spectra and mixings
is in how it affects the width. The determination of the effect of
the neutralino mixing on the width is straight forward when looking
at the Feynman rules for diagrams like those in Figure~\ref{fig:two}
and \ref{fig:four}. In both cases, the neutralino decay chain begins
with an off-shell squark, but only the wino and bino components of
the neutralino will couple to the squark. The smaller these components,
the smaller the total width will become. In the following we will
study in more detail the decay width for $\tilde{\chi}_{1}^{0}\to q\bar{q}\tilde{a}$.

Overall, we found that the KSVZ axino in the window of smaller $f_{a}$
has a rather predictive signal. The multiple displaced jets and missing
energy signature is not unique, but can at least in principal be distinguished
from the more well studied alternatives for neutralino decays and
the signal is not particularly sensitive to the choice of the PMSSM
benchmark model.

\subsection{The decay width for $\tilde{\chi}_{j}^{0}\to q\bar{q}\tilde{a}$}

\label{sec:width}

The neutralino decay width of Figure~\ref{fig:four} for massless
quarks and with universal squark masses $m_{\tilde{q}}=m_{\tilde{q}_{L}}=m_{\tilde{q}_{R}}$
is given by~\cite{Barnett:1987kn} 
\begin{eqnarray} \label{eq:width}
\Gamma(\tilde\chi_j^{0} \to q \bar q \tilde a) &=& \frac{m_{\chi^{0}_j} \alpha g_{\mathrm{eff}}^2}{64 \pi^2 \sin^2\theta_w} 
\frac{3}{2} \sum_q 16 [(T_{3q} Z_{j2}+(Q_q-T_{3q}) Z_{j1} \tan\theta_w)^2 + Q_q^2 Z_{j1}^2 \tan^2\theta_w] 
\nonumber \\ 
& \times &  [g(m_{\tilde a}^2/m_{\tilde \chi^{0}_j}^2,m_{\tilde q}^2/m_{\tilde \chi^{0}_j}^2)+
h(m_{\tilde a}^2/m_{\tilde \chi^{0}_j}^2,m_{\tilde q}^2/m_{\tilde \chi^{0}_j}^2)] \; ,
\end{eqnarray}
where $Z_{ji}$ are the matrix elements of the matrix which diagonalizes
the neutralino mass matrix, $\theta_{w}$ is the weak mixing angle,
and $Q_{q}=(2/3,-1/3),T_{3q}=(1/2,-1/2)$ for (up,down)-type quarks.
The effective coupling $g_{eff}$ is given by Eq.~\ref{eq:geff}
and the functions $g,h$ are provided in~\cite{Barnett:1987kn}.

It was emphasized in Section~\ref{sec:signal} that the displaced
multi-jet and MET signal was the only one that need be considered
for decays to axinos, but one can imagine that with the lightest neutralino
as a sufficiently pure Higgsino that diagrams like in Figures~\ref{fig:two}
and \ref{fig:four} would be suppressed enough that another decay channel
can dominate. While other axino channels can have a larger partial
width than the 2 and 3 jet channels for a very pure Higgsino, these
channels are still very much suppressed themselves, as they will contain
additional final-state particles or additional off-shell sparticles
or both. This possibility was explored for the lighter benchmark only
by varying the mixing parameters and retaining the same mass spectrum.
For this benchmark case the alternative channels, containing additional
gauge bosons or a Higgs boson only began to become competitive with
the multi-jet channels once the decay length was already several orders
of magnitude outside of the detector (hundreds of kilometers instead
of meters). Even though this possibility was only explored for a single
benchmark, it seems unlikely that any choice of spectra could reduce
the decay length by enough that it would matter to the phenomenology.
The effect of varying the Peccei-Quinn scale $f_{a}$ over the allowed
window is also relatively straight-forward (see Eq.~\ref{eq:width}).
How exactly the neutralino width scales with $f_{a}$ will depend
on which of the two couplings is dominant, but in either case the
width will vary by about two orders of magnitude.

\begin{figure}
\centerline{\includegraphics[width=4.5in]{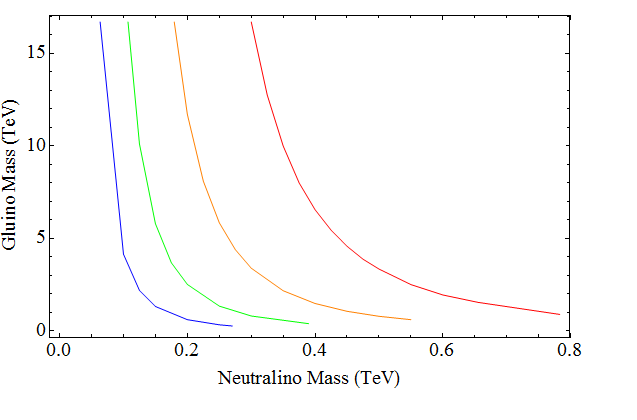}}
\caption{Contours of constant neutralino decay length $c\tau$ for decays to
an axino and two jets. Red is 0.01~m, yellow is 0.1~m, green is
1~m and blue is 10~m. All squarks are at 2 TeV and the axino is
taken to be massless. \label{fig:sixteen}}
\end{figure}

\begin{figure}
\centerline{\includegraphics[width=4.5in]{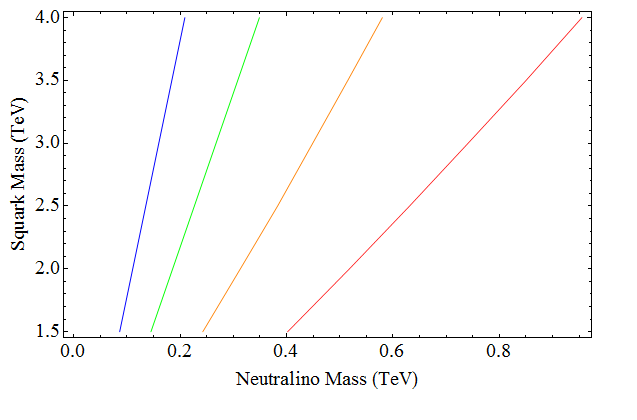}}
\caption{Contours of constant neutralino decay length $c\tau$ for decays to
an axino and two jets. Red is 0.01m, yellow is 0.1m, green is 1m and
blue is 10m. The gluino mass here is 3 TeV and the axino is taken
to be massless. \label{fig:seventeen}}
\end{figure}

\begin{figure}
  \centerline{\includegraphics[width=4.5in]{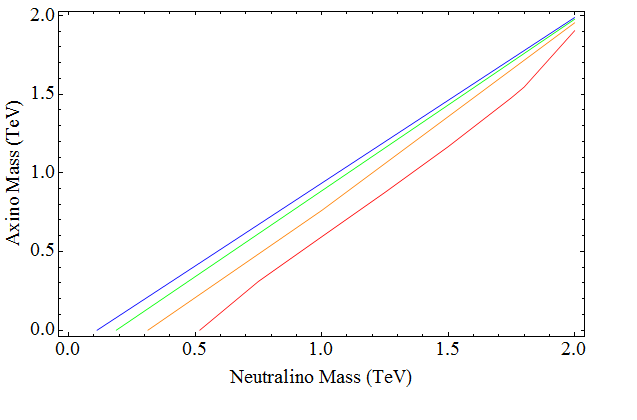}} \caption{Contours
    of constant neutralino decay length $c\tau$ for decays to an axino
    and two jets. Red is 0.01m, yellow is 0.1m, green is 1m and blue
    is 10m. The gluino mass here is 3 TeV and all squark masses are at
    2 TeV. \label{fig:eightteen}}
\end{figure}

\begin{figure}
\centerline{\includegraphics[width=4.5in]{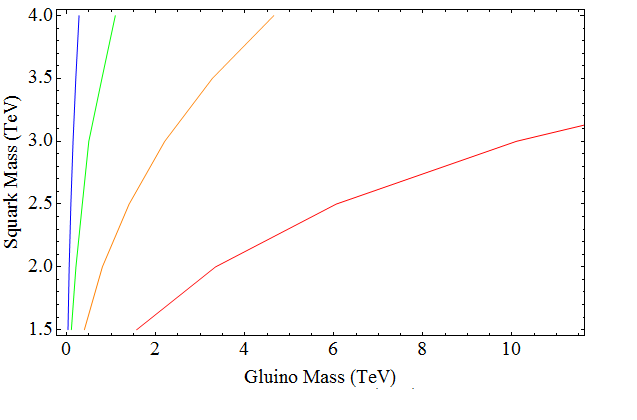}}
\caption{Contours of constant neutralino decay length $c\tau$ for decays to
an axino and two jets. Red is 0.01m, yellow is 0.1m, green is 1m and
blue is 10m. The neutralino mass here is 0.5TeV and the axino is taken
to be massless. \label{fig:nineteen}}
\end{figure}

The other parameters affecting the width are all sparticle masses:
The axino mass, the neutralino mass, the squark mass and the gluino
mass. In Section~\ref{sec:signal} it was stated that several of
the Snowmass PMSSM benchmarks from the collection in \cite{Cahill-Rowley:2013gca}
allowed for neutralino decays to axinos with a decay length appropriate
for searches at the LHC, even though only two were chosen for simulation,
and it seems as though this scenario could be rather common for SUSY
models with sparticle masses in the range that is explorable in the
near future. With the decay width of neutralinos to axinos depending
on so many variables it is difficult to bound exactly what the model
space is available to such searches, but Figures~\ref{fig:sixteen}
through ~\ref{fig:nineteen} make an attempt of demonstrating what
range of SUSY parameters would allow for this type of signal. Each
is a plane in parameter space that shows contours of equal neutralino
decay length $c\tau$ for the 2 jet plus axino signal only, as given
in Eq.~\ref{eq:width}. For very heavy neutralinos the channel to
heavy quarks opens up, and for squarks much heavier than gluinos the
3 jet signal will start to become competitive and eventually dominate.
In each of these plots the neutralino is taken to be a very pure bino
and the Peccei-Quinn scale $f_{a}$ takes its lowest value in the
allowed window, so that these planes of parameters space are already
at there ``least displaced'' for these parameters. These plots also
show for what SUSY masses prompt decays to axinos may be possible,
though this signal comes with its own set of challenges that are not
discussed here.

A compressed spectrum (with a neutralino NLSP mass close to the gluino
mass) will make the off-shell decay to axinos easier, resulting in
a shorter mean decay length. The spectra can become very compressed
and the decay will still be displaced, (especially for larger values
of $f_{a}$), perhaps providing an easier discovery channel than otherwise
available for a compressed spectrum. For less compressed spectra,
or spectra with a lighter neutralino, the mean decay length increases.
When considering a larger parameter space of SUSY models (as the PMSSM
does), compressed spectra are not uncommon \cite{LeCompte:2011cn}.
While more compressed spectra can be very difficult to search for
at colliders \cite{Dreiner:2012gx}, in this case the primary effect
on the signal is on the total width of the NLSP, and typically more
compressed spectra will simply have the displaced jets closer to the
primary vertex (with some fraction still being more displaced). This
is an interesting scenario in and of itself, implying that an otherwise
difficult to study spectrum at the LHC may have axino production as
its discovery channel.

\section{Dark  Matter abundances in the hadronic axion window}

\label{sec:remarks}

There are also many unanswered questions concerning the cosmology of
such a model. In this work, we only attempt to argue that there are
enough parameters that can be adjusted and that such a model has all
the right ingredients for a working cosmology, but this does not
guarantee such a cosmology exists. It seems as though a low reheating
temperature ($T_{rh}$) is required to make this scenario viable. With
a reheating temperature lower than the freeze out temperature ($T_F$)
for axions, constraints from large scale structure and CMB
measurements can be avoided and the hadronic axion window is still
viable \cite{Grin:2007yg}.  The size of axion and axino DM abundances
depends on a number of factors that have not been specified here
because they do not effect the collider signal. The phenomenology here
is relatively insensitive to the axino mass, which will effect the
size of its abundance and how relativistic it is. Late decays of the
saxion can effect the size of both the axion and axino population, and
can inject extra entropy into the early universe to dilute these
species, so the role of the saxion is non trivial and requires further
study in this scenario.  The gravitino was assumed to be heavy enough
not to effect the collider phenomenology in this scenario, but it too
could play a more complicated role. A light enough gravitino can have
a comparable coupling with the LOSP as the axino (as shown in
Section~\ref{sec:results}), but can also be coupled more strongly or
weakly depending on its mass.  While only an LSP gravitino is likely
to impact the collider phenomenology discussed here, an intermediate
mass gravitino can still effect the cosmology with late decays to
other states. Whether or not there are RPV couplings present can also
effect both the axino and axion abundance.  There are several options
for what types of RPV couplings there are (if any) and the size of
each coupling has a wide allowed range. The axion/axino cosmology is
more sensitive to the choice of RPV couplings in this case, because
the axion/axino couplings are restricted in the hadronic axion window,
and so there are not as many options for decay chains.

With so many possible variables, we only attempt to illustrate the
viability of such a scenario in the simplest case, i.e. with no RPV
and ignoring the possible effects of gravitinos and saxions. In this
case there are still several possible populations of DM: thermal and
non-thermal relics from both axions and axinos.

For $f_a>T_{rh}$ the contribution from non-thermal axions to the cold
relic abundance from the misalignment mechanism is approximated
by~\cite{Beltran:2006sq} (for a review see, e.g.,
\cite{Steffen:2008qp}):
\begin{equation}\label{eq:nonthermal}
\Omega_{a}^{NTP}h^{2}\approx0.15\xi f(\theta_i^2) \theta_i^2 \left(\frac{f_{a}}{10^{12}GeV}\right)^{7/6} \; 
\end{equation}
with $\xi={\cal O}(1)$ parameterizing theoretical uncertainties,
$\theta_i$ denoting the misalignment angle with respect to the
CP-conserving position, and $f(\theta_i^2)$ accounts for anharmonic
corrections due to large values of $\theta_i$ ($f(\theta_i^2) \approx
1.2$~\cite{Beltran:2006sq}). Even when assuming $|\theta_i|=\pi$ this
contribution from non-thermal axions to the CDM abundance is negligible
for $f_a$ in the hadronic window.

\begin{figure}
  \centerline{\includegraphics[width=4.5in]{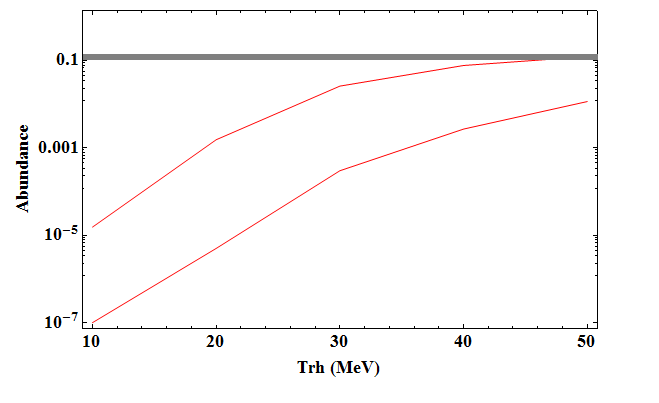}} \caption{The
    thermal relic abundance of axions as a function of reheating
    temperature. Two curves are shown, as the scale $f_{a}$ (or
    equivalently the axion mass $m_{a}$) is varied over its possible
    values within the hadronic axion window.  The approximate measured
    value of the total dark matter relic abundance is shown by the
    solid gray line ($\Omega_{DM}h^{2}=0.119$)\cite{Planck:2015xua}.
 \label{fig:twenty}}
\end{figure}

The thermal axion abundance at very low reheating temperatures is
calculated numerically in~\cite{Grin:2007yg}. Using that the axion
mass $m_a$ is related to the PQ scale $f_a$ as~\cite{Agashe:2014kda}
\begin{equation}\label{eq:fama}
f_a = \frac{\sqrt{z}}{1+z} \frac{f_\pi m_\pi}{m_a} \approx \frac{6 \times 10^{6} {\rm eV} \, {\rm GeV}}{m_a (\rm{in} \, \rm{eV})}
\end{equation}
we use the results for $\Omega_{a}/\Omega_0(T_{rh})$ provided in
Figure~4 of \cite{Grin:2007yg} for $5 \rm{eV} \le m_a\le 20$~eV
(corresponding to $3 \times 10^5 {\rm GeV }\le f_a \le 1.2 \times
10^{6}$ GeV) to estimate $\Omega_{a}h^2$ for the hadronic window for
$f_a$. We assume $\Omega_{0}h^2=m_a/13 {\rm eV}/g_{*_S,F}$ with the
effective number of relativistic degrees of freedom $g_{*_S,F} \approx
10$ at freeze-out temperature $10 \, {\rm MeV} <T_F<100$
MeV~\cite{Hannestad:2005df}.  The resulting thermal axion abundance as
a function of reheating temperature $T_{rh}$ is plotted in red in
Figure \ref{fig:twenty} for $f_{a}=3 \times 10^5 {\rm GeV }$ and $f_a
=1.2 \times 10^{6}$ GeV. Axion constraints are completely lifted in
the lower range of $T_{rh}$ shown, but rising towards 50 MeV, the
constraints may become relevant again with a dependence on $f_{a}$
(see \cite{Grin:2007yg}).

\begin{figure}
  \centerline{\includegraphics[width=4.5in]{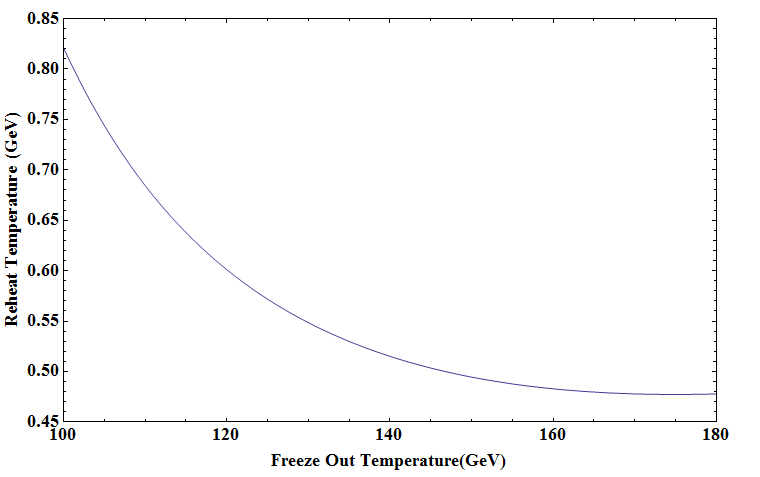}} \caption{The
    reheating temperature vs the freeze out temperature for the scenario
    of thermal axino production at low reheating temperature for an
    example case at $f_{a}=3 \times 10^5 {\rm GeV }$ with squarks and gluinos with masses 1 and 2 TeV
    respectively. At lower reheating temperatures axinos will freeze out
    sooner, and be diluted more severely.
 \label{fig:twenty_one}}
\end{figure}

The other possible major component expected is the thermal abundance
of axinos. With a lower $f_{a}$ scale one may expect the thermal
axinos to over-close the universe, but they too will be diluted due to
a lower reheating temperature.  An approximation of this dilution is
given in \cite{Roszkowski:2014lga}, which may be used to rescale the
abundance at high $T_{rh}$ down to its value in the low reheating
scenario:
\begin{equation}\label{eq:omegalow}
\Omega_{DM}h^{2}\thickapprox\left(\frac{T_{rh}}{m_{\chi}}\right)^{3}\left(\frac{m_{\chi}}{T_{F}}\right)^{3}\Omega_{DM}h^{2}(\mathrm{High }T_{rh}) \, .
\end{equation}
The thermal axino abundance for a sufficiently large reheating
temperature is given by \cite{Rajagopal:1990yx}:
\begin{equation}\label{eq:omegalarge}
\Omega_{\tilde{a}}h^{2}=\frac{m_{\tilde{a}}}{2 KeV} \, .
\end{equation} 
Thermal axino production at low reheating temperatures was explored in
more detail in \cite{Roszkowski:2015psa}, but only down to reheating
temperatures as low as 1~GeV. At even lower temperatures axinos become
decoupled from kinetic equilibrium but still remain in chemical
equilibrium.  For a treatment of this scenario, we follow the
methodology of \cite{Monteux:2015qqa}, which originally studied
goldstino production at low reheating temperature, but can be applied
to axinos by adjusting the corresponding couplings. At lower reheating
temperatures axinos freeze out earlier, and will be diluted more
greatly by entropy produced during inflation. Assuming goldstino
production via squark decay described by $\Gamma_{\tilde q \to \xi
  q}=m_{\tilde q}^5/(16 \pi^2 F_\xi^2)$, the freeze out temperature in
this scenario is determined in \cite{Monteux:2015qqa} from:
\begin{equation}\label{eq:freezeout}
3H(T_{F})=1.4(5\pi^{2}g_{*}/72)^{1/2}\frac{T_{F}^{4}}{M_{P}T_{R}^{2}}\simeq\frac{12m_{\tilde{q}}^{5}}{16\pi F_\xi^2}\sqrt{\pi}\left(\frac{m_{\tilde{q}}}{T_{F}}\right)^{3/2}e^{-m_{\tilde{q}}/T_{F}} \, .
\end{equation}
In \cite{Covi:2002vw} the decay width for axino production via $\tilde
q \to q \tilde a$ is given as $\Gamma_{\tilde q \to \tilde{a}
  q}=g_{eff}^2 m_{\tilde q}/(16 \pi^2)$ with $g_{eff}$ of
Eq.~\ref{eq:geff}, so that the freeze out temperature can be
determined from Eq.~\ref{eq:freezeout} by replacing $F_\xi$
accordingly:
\begin{equation}\label{eq:trafo}
F_\xi \to \frac{m_{\tilde q}^2}{g_{eff}} \, .
\end{equation}
The resulting dependence of the reheating temperature on the freeze
out temperature is shown in Fig.~\ref{fig:twenty_one}, for an example
case at $f_{a}=3 \times 10^5 {\rm GeV }$ with squarks and gluinos with
masses 1 and 2 TeV respectively. As the reheating temperature
approaches the low values necessary to avoid axion constraints, the
freeze out temperature gets larger very quickly, and it is the ratio
of these two scales which will determine how diluted the thermal
abundance is.  Using the reheating and freeze out temperatures from
Fig.~\ref{fig:twenty_one} and the more accurate method of
\cite{Monteux:2015qqa} with $F_\xi$ adjusted as in Eq.~\ref{eq:trafo},
we find the relic abundance to be many orders of magnitude below the
amount observed today for the range of axion masses considered here
($\Omega_{\tilde a}h^2 \approx 10^{-5} m_{\tilde a}/{\rm GeV}$). To be
able to obtain the abundance for $T_{rh}<100$~MeV from the method used
in~\cite{Monteux:2015qqa} would require a re-evaluation of the approximations used
in the calculation, which is beyond the scope of this work, but the
expected effect of going to even lower reheat temperature is to have
a thermal axino abundance even smaller than this already negligible
amount.

There is also the possibility of a non-thermal axino abundance, from
the decays of neutralinos, but its number density is effectively
diluted twice. Before considering the diluting effects of going to a
very low reheating temperature, the number density of non-thermal
axinos is inherited from neutralinos but scaled by the ratio of
masses,
\begin{equation}
\Omega_{\tilde{a}}^{NTP}h^{2}=\frac{m_{\tilde{a}}}{m_{\tilde{\chi}_{1}^{0}}}\Omega_{\tilde{\chi}_{1}^{0}}h^{2}.
\end{equation}
For a reasonable range of axino masses (between MeV and a few GeV),
this already results in a large reduction of the non-thermal axino
abundance.  On top of this, $\Omega_{\tilde{\chi}_{1}^{0}}h^{2}$ must
be rescaled for a very low reheating temperature, since the neutralino
number density itself would also have been diluted away during
inflation.  Even for SUSY spectra that grossly overproduce bino NLSPs
at high $T_{rh}$, ($\Omega_{\tilde{\chi}_{1}^{0}}h^{2}(\mathrm{High
}T_{rh}) >10^5$), the neutralino relic abundance (and by extension the
non-thermal axino abundance) is usually negligible for reheating
temperatures small enough to avoid axion constraints ($T_{rh}<100$
MeV). Only when both the neutralino abundance and the axino mass are
very large, the non-thermal axino component may be sizable.  Axinos in
this mass range (approaching the neutralino mass) are unlike those in
the benchmarks whose phenomenology is studied here (where axinos were
considered light), but heavier axinos should be easier to distinguish
from similar decays with gravitinos or with RPV.  The most likely
scenario then seems to be a dark matter abundance that is mostly
thermally produced axions, but with the possibility of a significant
fraction being non-thermal axinos, if they are heavy enough (compared
to the neutralino mass). If these two populations do not saturate the
relic abundance there is also the possibility of inflaton decays
directly to axinos, which would provide an additional component, but
this is highly model dependent.  This is just one possible scenario,
without considering gravitinos, saxions or RPV, but it should evade
all existing constraints and such a model will have a collider signal
that is very similar to the signal discussed in this work.

\section{Conclusions}
\label{sec:conclusion}

Supersymmetric models with axions and axinos are very attractive extensions
of the SM since they can address issues of naturalness in QCD, in
the electroweak sector, and with regards to dark matter. The one feature
of these types of models that could be considered disappointing is
that the additional particles, the axion and the axino, can be rather
difficult to detect. The scenario proposed here is a PQMSSM model with
a light LSP axino with only QCD couplings and a neutralino NLSP. The
signal studied here is the production of neutralinos and their displaced
decay to two jets and a KSVZ axino via an effective squark-quark-axino
coupling. This scenario could be detectable at the 14 TeV LHC provided
the Peccei-Quinn scale can exist in the smaller range $3\times10^{5}\,{\rm GeV}<f_{a}<3\times10^{6}\,{\rm GeV}$
(hadronic axino window). We did not consider sneutrino NLSPs, in which
case the topologies for NLSP decays becomes more varied and can include
photons and charged leptons, making the phenomenology more complicated,
especially in distinguishing from RPV and gravitino scenarios.

The scenario of the hadronic axion window is not new, and its cosmology
has been discussed in the literature (see, e.g.,~\cite{Moroi:1998qs})
but the consequences of having this window in a SUSY model have not
been explored until recently, and there is still much to
learn. This is not the only scenario that allows axinos to be detectable
at colliders, but to the authors' best knowledge it is the only way
currently proposed to detect KSVZ axinos with a neutral NLSP. This
scenario gives a predictive collider signature due to its limited
couplings. We have shown that it has the potential to be distinguishable
from similar models with neutralino decay, and that this signature
is relatively insensitive to the choices of MSSM parameters. While
we find that there is potential for the LHC to be sensitive to the
scenario studied here, a detailed detector simulation that implements
for instance the triggers used in hidden valley searches~\cite{Aad:2013txa}
is needed to fully assess its observability.

It is interesting to probe the hadronic axion window via collider
searches for a variety of reasons. While it has been argued extensively
in the literature that there can be a variety of benefits to having
SUSY with axions, there are very few ways to test the axion coupling
$f_{a}$ independent of its photon or electron coupling, which this
scenario allows for. While there is still much to learn about this
scenario, there are tentative hints that it could have attractive
features beyond a detectable axino. It may also provide a discovery
channel for otherwise difficult to study compressed SUSY spectra,
it may alleviate some issues of tuning, and the cosmology it fits in
may have other interesting consequences, such as detectable RPV decays that are
competitive with decays to axinos. This scenario is still very new,
both for collider studies and for cosmology, and much more work is
required to determine its viability, detectability and consequences.

\begin{acknowledgments}
  We thank H.~Baer for valuable discussions and comments, and for
  reading an early version of the manuscript.  We also would like to
  thank Ian Woo Kim for his assistance with evchain and Benjamin Fuks
  for his assistance with FeynRules. The work of C.S.R. is supported
  in part by the National Science Foundation under award
  no.~PHY-1118138 and a LHC Theory Initiative Graduate Fellowship, NSF
  Grant No.~PHY-0705682.  The work of D.W. is supported in part by the
  National Science Foundation under award no.~PHY-1118138.
\end{acknowledgments}

\bibliography{main}

\end{document}